\documentclass[twocolumn,amsmath,amssymb,aps]{revtex4-2}
\usepackage{lineno,hyperref}
\usepackage{bm}
\usepackage{amsmath}
\usepackage{gensymb}
\usepackage{epstopdf}
\usepackage{booktabs}

\usepackage{graphicx}
\usepackage{ulem}
\usepackage{xcolor}
\begin{document}

\title{Coexisting structural disorder and robust spin-polarization in half-metallic FeMnVAl}

\author{Shuvankar Gupta$^1$, Sudip Chakraborty$^1$, Santanu Pakhira$^2$, Celine Barreteau$^3$, Jean-Claude Crivello$^3$,  Bilwadal Bandyopadhyay$^1$, Jean Marc Greneche$^4$, Eric Alleno$^3$, and Chandan Mazumdar$^1$ }
\email{chandan.mazumdar@saha.ac.in}
\affiliation{$^1$Condensed Matter Physics Division, Saha Institute of Nuclear Physics, A CI of Homi Bhabha National Institute, 1/AF, Bidhannagar, Kolkata 700064, India}
\affiliation{$^2$Ames National Laboratory, Iowa State University, Ames, Iowa 50011, USA}
\affiliation{$^3$Univ. Paris-Est Creteil, Institut de Chimie et des Mat\'{e}riaux Paris-Est, UMR 7182 CNRS UPEC, 2 rue H. Dunant, 94320 Thiais, France}
\affiliation{$^4$Institut des Mol\'{e}cules et Mat\'{e}riaux du Mans, IMMM, UMR CNRS 6283, Le Mans Universit\'{e}, Avenue Olivier Messiaen, Le Mans Cedex 9, 72085, France}

\date{\today}

\begin{abstract}

Half-metallic ferromagnets (HMF) are on one of the most promising materials in the field of spintronics due to their unique band structure consisting of one spin sub-band having metallic characteristics along with another sub-band with semiconductor-like behavior. In this work, we report the synthesis of a novel quaternary Heusler alloy FeMnVAl and have studied the structural, magnetic, transport, and electronic properties complemented with first-principles calculations. Among different possible structurally ordered arrangements, the optimal structure is identified by theoretical energy minimization. The corresponding spin-polarized band structure calculations indicates the presence of a half-metallic ferromagnetic ground state. A detailed and careful investigation of the x-ray diffraction data, M\"{o}ssbauer and nuclear magnetic resonance spectra suggest the presence of site-disorder between the Fe and Mn atoms in the stable ordered structure of the system. The magnetic susceptibility measurement clearly establishes a ferromagnetic-like transition below $\sim$213 K. The ${^{57}}$Fe M\"{o}ssbauer spectrometry measurements suggest only the Mn-spins could be responsible for the magnetic order, which is consistent with our theoretical calculation. Surprisingly, the density-functional-theory calculations reveal that the spin-polarization value is almost immunized (92.4\% ${\rightarrow}$ 90.4\%) from the Mn-Fe structural disorder, even when nonmagnetic Fe and moment carrying Mn sites are entangled inseparably. Robustness of spin polarization and half metallicity in the studied FeMnVAl compound comprising structural disorder is thus quite interesting and could provide a new direction to investigate and understand the exact role of disorders on spin polarization in these class of materials, over the available knowledge.

\end{abstract}

\maketitle

\section{\label{sec:Introduction}Introduction}

In recent years, research on spintronics and related materials have emerged as one of the most exciting and promising branch in the field of magnetic materials, material sciences, condensed matter physics and magneto-electronic devices~\cite {vzutic2004spintronics}. Unlike contemporary electronics, which is related to charge of the carrier particles, spintronics deals with spin of electron and the associated magnetic moment~\cite{wolf2001spintronics}. Specialized semiconductor materials are needed in electronics in order to regulate the passage of charge through transistors. Since generating a current to maintain electron charges in a device is more energy-intensive than changing spin, spintronics devices are more energy-efficient. Data transmission is expedited by the ease with which spin states can be changed. As spin is non-volatile because the spin of electron is not energy-dependent, the information transmitted through spin remains stable even when energy is lost ~\cite{vzutic2004spintronics,wolf2001spintronics,felser2007spintronics}. Materials having higher spin-polarization are generally considered ideal for application in spintronics devices~\cite{felser2007spintronics}.

Half-metallic ferromagnets (HMF) are one such promising materials, often known to exhibit high spin-polarization~\cite{katsnelson2008half}. The band structure of HMF are comprised of one spin sub-band that shows metallic behavior and another spin sub-band with semiconductor-like behaviour, resulting in a unique band mechanism capable of very high spin-polarization~\cite{de1983new}. Theoretically, even 100 \% spin-polarized current can be achieved in HMF. Among different classes of reported  HMFs, Heusler alloys have attracted extensive attention in the research direction of spintronics and related phenomena due to their high Curie-temperature ($T_{\rm C}$) and tunable electronic structure~\cite{wurmehl2005geometric, shan2009demonstration, graf2011simple}.

Full Heusler alloys, belonging to Cu$_2$MnAl-type, are represented stoichiometrically by $X_2YZ$ (where $X$, $Y$ are the transition elements and $Z$ is the \textit{s-p} group elements) and are found to crystallize in $L2_1$-type structure where $X$, $Y$ and $Z$ atoms occupy $8c$, $4b$ and $4a$ sites respectively in space group $Fm\bar{3}m$ (No. 225)~\cite{graf2011simple}. Such crystal structure consists of four interpenetrating face-centered-cubic (\textit{fcc}) sublattices. Interestingly in $X_{2}YZ$, if one of the $X$ atoms is replaced by a different transition element {$X'$}, a quaternary Heusler alloy with a crystal structure of the Y-type (LiMgPdSn-type) is formed, in which four interpenetrating sublattices are formed with four different atoms~\cite{bainsla2016equiatomic}. As a consequence, the $8c$ site in $Fm\bar{3}m$ is split in $4c$ and $4d$ in space group: $F\bar{4}3m$ (No. 216). From the point of view of both fundamental physics and application oriented research, quaternary Heusler alloys occupy an important place, as some of them exhibit the recently discovered characteristics of spin-gapless semiconductors (SGS)~\cite{bainsla2016equiatomic, bainsla2015spin, bainsla2015origin,venkateswara2018competing}.

SGS is a subclass of HMF materials in which one spin sub-band is semiconducting and another spin sub-band is semi-metallic ~\cite{wang2008proposal, ouardi2013realization}. In the presence of such a unique band structure, SGSs are not only capable of yielding 100 \% spin- polarization but also exhibit very high electronic mobility ~\cite{ouardi2013realization}. Furthermore, with the application of external perturbations (electric field, pressure, magnetic field), one can easily tune and switch between $n$- and $p$-type spin-polarized carriers; making these materials perfect candidate for spintronics application~\cite{bainsla2016equiatomic}. So far, only a limited number of such quaternary Heusler alloys are reported and thus the advancement of this field of research strongly depends on the discovery of new materials of this type. Among the experimentally reported members, only a few Co-based  and Ni-based quaternary Heusler alloys are explored so far~\cite{bainsla2016equiatomic, bainsla2015spin, rani2017structural, samanta2020structural}. Research on Fe-based quaternary Heusler alloys is in its infancy~\cite{venkateswara2019coexistence}, although some theoretical predictions exist \cite{khandy2019lattice,shakil2021determination,amudhavalli2017first}. Furthermore, the presence of structural disorder in various Heusler-based HMFs is quite inherent and it is almost impossible to obtain a disorder-free crystal structure. The presence of disorder can have a substantial impact on the observed spin-polarization value~\cite{miura2004atomic}. Thus, in addition to synthesizing new HMF with the aim of having negligible/mimimum amount of disorder, it is also of great interest to probe the disorder present in the material and to examine its role on the observed  physical, transport, and spin- polarization value and related phenomena.

In this work, we synthesized the new quaternary Heusler alloy FeMnVAl and experimentally explored its structural and physical properties complemented by band structure calculations using density functional theory (DFT). The presence of microscopic local structural disorders was also identified by two different nuclear magnetic spectrometry techniques, ${\textit{viz}}$.,  M\"{o}ssbauer spectrometry and nuclear magnetic resonance (NMR). The magnetic and transport properties of this system have also been probed and discussed using different experimental techniques and theoretical investigations.

\section{\label{sec:Methods}METHODS}
\subsection{Experimental}
The polycrystalline FeMnVAl was synthesized by arc melting process taking appropriate high purity ($>$99.9 \%) constituent elements under an argon atmosphere. The sample was melted 5-6 times, turning it over after each melt to achieve better homogeneity. To compensate the amount of Mn evaporated, an additional 2\% extra Mn was added during the melting. For distinguishing some elemental contributions in NMR spectra, the compound FeMnVAl$_{0.5}$Ga$_{0.5}$ was also  synthesised as a reference material. Room temperature  powder X-ray diffraction (XRD) measurements were carried out using Cu-K$\alpha$ radiation on TTRAX-III diffractometer (Rigaku, Japan). The single-phase nature  and crystal structure of the sample were determined  by performing  a Rietveld refinement using the FULLPROF software package~\cite{rodriguez1993recent}. Magnetic measurements were performed in a commercial SQUID-VSM  (Quantum design Inc.(USA)) in the temperature range 3--380 K and magnetic fields up to 70 kOe\@.  Resistivity measurements have been carried out using Physical Property Measurement System (Quantum design Inc. (USA)) with the standard four-probe technique. The hyperfine structures of the Fe sites were studied by ${^{57}}$Fe transmission M\"{o}ssbauer  spectrometry to get information on the local environment of ${^{57}}$Fe at the atomic scale. Spectra were obtained at 300 K and 77 K using an electromagnetic transducer with a triangular velocity form and a ${^{57}}$Co source diffused into a Rh matrix and a bath cryostat. The samples consist of a thin layer of powder containing about 5 mg Fe/cm$^2$. The hyperfine structures were modelled by means of a least-square fitting procedure involving quadrupolar doublets and magnetic sextets composed of Lorentzian lines using the in-house program `MOSFIT'. The values of isomer shift are quoted to that of ${\alpha}$--Fe at 300 K while the velocity was controlled by using a standard of ${\alpha}$--Fe foil. Nuclear magnetic resonance (NMR) measurements were performed with a Bruker 300 MHz Ultrashield magnet operated at 70.045 kOe, Thamway PROT4103MR spectrometer and a cryostat of Oxford Instruments. The cryostat houses the home-made NMR sample rod that has `tuning' and `matching' capacitors, and a Cernox temperature sensor near the $\textit{rf}$ sample coil. Spectra at different temperatures in the range of  80-295 K were recorded by the process of integration of the spin-echo signals while sweeping the $\textit{rf}$ in discrete steps over the position of resonance using a $\frac{\pi}{2}-\tau-\frac{\pi}{2}-\it{solid} \it{echo}$ pulse sequence.

\subsection{Computational}	
First principles calculations in the frame of DFT were conducted using the projector augmented wave (PAW) method ~\cite{blochl1994projector} implemented in the Vienna \textit{ab initio} simulation package (VASP)~\cite{kresse1993ab, kresse1994norm}. The exchange-correlation was described by the generalized gradient approximation modified by Perdew, Burke and Ernzerhof (GGA-PBE)~\cite{perdew1996generalized}.  Plane waves with a cutoff of E = 600\,eV were included in all calculations. After performing the volume and ionic (for disordered compounds) relaxation steps, the tetrahedron method with Bl\"{o}chl correction~\cite{blochl1994improved} was applied. Spin-polarization calculations were systematically implemented. In order to model statistical chemical disorder in FeMnVAl, unit cells based on the concept of special quasirandom structure (SQS)~\cite{zunger1990special} were used. To generate the SQS, the cluster expansion formalism for the multicomponent and multisublattice systems~\cite{sanchez1984generalized} was used as implemented in the Monte-Carlo code (MCSQS) containing the Alloy-Theoretical Automated Toolkit (ATAT)~\cite{van2009multicomponent, van2013efficient}. Subsequent DFT calculations were performed to test the quality of the SQS and to see how reliable the DFT results were. The root mean square error (rms) was used as another quality criterion besides the calculations including a different order of interactions. The rms error describes the deviation of the correlation function of the SQS ($\Pi^{k}_{SQS}$) from the correlation function of a fully random structure ($\Pi^{k}_{md}$) for all k clusters.
\begin{equation}
\ rms = \sqrt{\sum_{k}(\Pi^{k}_{SQS}-\Pi^{k}_{md})^2}
\label{eq1}
\end{equation}
\indent
Several tests on the dependence of the type and number of clusters were performed to generate the disordered structure (Al at 4$a$ (0,0,0), V at 4$b$ (0.5,0.5,0.5) and Fe=0.5/Mn=0.5 at 4$c$ (0.25,0.25,0.25) and Fe=0.5/Mn=0.5 at 4$d$ (0.75,0.75,0.75). Finally, 7 pairs, 5 triplets and 11 quadruplets interactions were considered to obtain reliable results (see the distribution of the \textit{k}-chosen clusters in Supplementary Material~\cite{[{See Supplemental Material at }][{ for details.}]supp}).  For the fully disordered LiMgPdSn phase, a quaternary SQS cell of 28 atoms was generated. The heat of formation $\Delta_f{H}$ has been calculated by total energy difference with pure elements in their stable ground state (\textit{i.e.} ferromagnetic bcc Fe, \textit{etc}).

\section{Results and Discussion}

\subsection{\label{sec:DOS1}Electronic structure calculations - Ordered structure}

\begin{figure}[ht]
\centerline{\includegraphics[width=.48\textwidth]{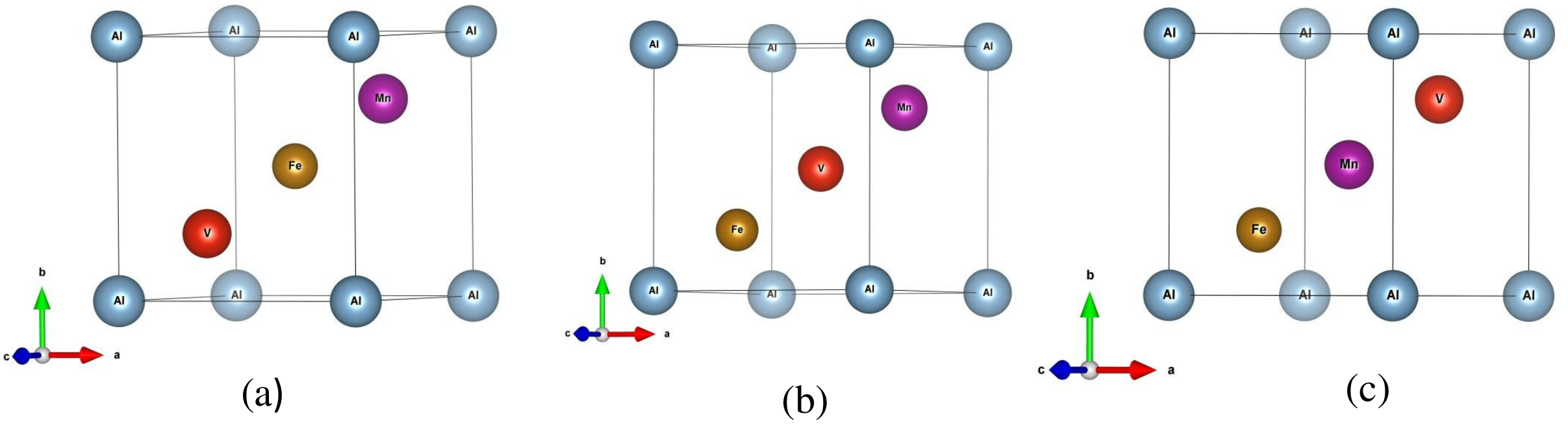}}
{\caption{Primitive unit cell representation of (a) Type 1 (b) Type 2 and (c) Type 3 ordered structure as described in Table~\ref{Enthalpy}. Color representations of atoms, Fe: brown-yellow ball, Mn: magenta ball, V: red ball and Al: light-blue ball.}\label{All_Structure}}
\end{figure}
In order to optimize the crystal structure and find the most stable configuration, density functional theory (DFT) calculations on FeMnVAl in the LiMgPdSn-type structure were first performed. In a quaternary Heusler alloy $XX'YZ$, if the $Z$ atoms are considered at position 4$a$ (0,0,0), the remaining three atoms $X$, $X'$ and $Y$ could be placed in three different fcc sublattices, namely, 4$b$ (0.5,0.5,0.5),  4$c$ (0.25,0.25,0.25) and 4$d$ (0.75,0.75,0.75). As the
permutation of the atoms in 4$c$ and 4$d$ positions results in energetically invariant configurations, out of a total six possible combinations, only three independent structures are feasible, they are represented in Fig.~\ref{All_Structure}. We considered these three configurations (Type-1,2 and 3) in our calculations and the results are summarized in Table~\ref{Enthalpy}.\\

\begin{table}[]
\caption{Calculated enthalpy of formation $\Delta_f{H}$ for each ordered structure type of FeMnVAl, and one disordered case (see text).}
\begin{tabular}{|c|c|c|c|c|c|}
\hline
 & 4\textit{a} & 4\textit{b} & 4\textit{c} & 4\textit{d} & $\Delta_f{H}$ (kJ/mol) \\ \hline
Type 1   & Al & Fe & Mn & V  & -2.31                            \\ \hline
Type 2   & Al & V  & Mn & Fe & -32.70                           \\ \hline
Type 3   & Al & Mn & V  & Fe & -0.96                            \\ \hline\hline
disordered & Al & V & Fe:Mn  & Fe:Mn & -34.14  \\
\hline
\end{tabular}
\label{Enthalpy}
\end{table}

\begin{figure*}[t]
\centerline{\includegraphics[width=.96\textwidth]{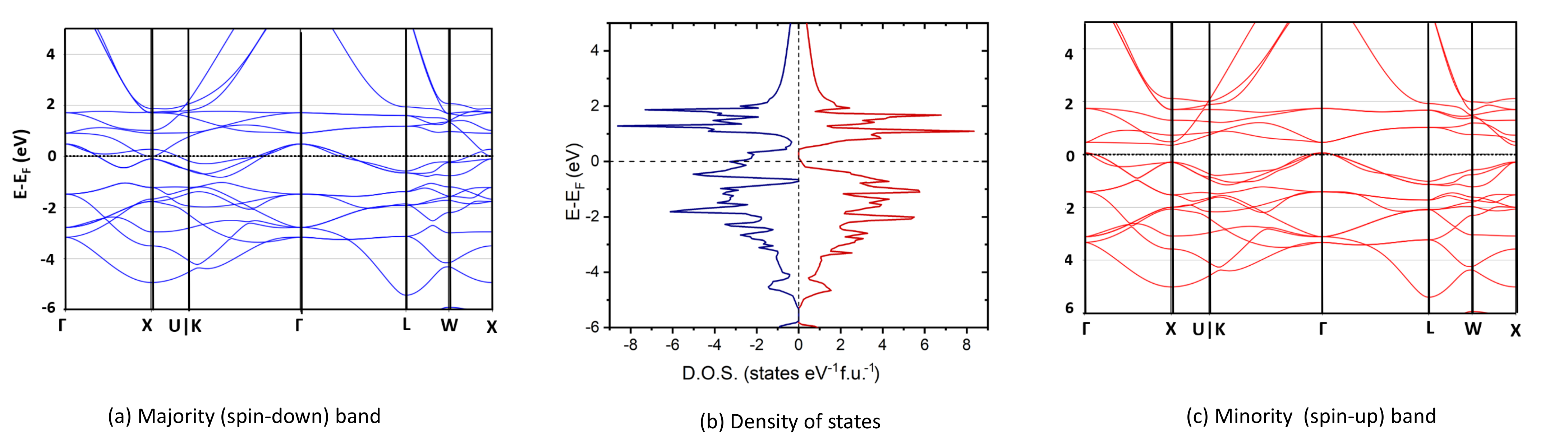}}
{\caption{Spin-polarized band structure and density of states of FeMnVAl in ordered Type 2 structure: (a) majority (spin-down) band (b) density of states, (c) minority (spin-up) band. The energy axis zero point has been set at the Fermi level, and the spin-up (minority) and spin-down (majority) electrons are represented by positive and negative values of the DOS, respectively.}\label{Ordered_DOS}}
\end{figure*}

According to our DFT calculations, Type 2 ordered structure is the most stable configuration yielding Al+V and Fe+Mn is the same cubic planes (100), respectively (Table~\ref{Enthalpy}).
This result is consistent with other quaternary Heusler compounds, where the least electronegative atom occupies 4\textit{b} position~\cite{graf2011simple} . Fig.~\ref{Ordered_DOS} shows the calculated spin-polarized band structure and  density of states (DOS) of the energetically most favorable configuration (Type 2).

Full and quaternary Heusler alloys in ordered configurations obey the Slater-Pauling (S-P) rule, which states that the total magnetic moment for Heusler alloys is governed by the relation
m=($N_V$-24) $\mu_{\rm B}$/f.u.,
where $N_V$ is the total valence electrons count (VEC) for a material~\cite{galanakis2002slater}. For transition metal-based alloys, $N_V$ is the number of outer ($s+d$) electrons for the main-group element, while it is the total number of outer ($s+p$) electrons for \textit{sp}-group element. The S-P rule is generally considered to be a very important criterion that relates the magnetism (the total spin-magnetic moment) to the electronic structure of a material, as all the reported half-metallic Heusler ferromagnets are known to obey the S-P rule~\cite{graf2011simple,bainsla2016equiatomic,galanakis2002slater}. In case of FeMnVAl, the total VEC is 23, and therefore the total magnetic moment should be -1 $\mu_{\rm B}$/f.u. Before presenting our results, it is important to point out that this compounds with VEC $<$ 24 has negative total spin moments and the gap is located at the spin-up band due to the S-P rule. Moreover, in contrast to the other Heusler alloys~\cite{galanakis2002slater}, the spin-up electrons correspond to the minority-spin electrons and the spin-down electrons to the majority electrons~\cite{galanakis2007doping}. From the DFT calculation, the total magnetic moment is estimated to be -0.97\,$\mu_{\rm B}$/f.u. which is in close agreement with the Slater-Pauling (S-P) rule. For the most stable configuration (Type~2), the magnetic moments  of the elements were also been estimated by the calculations at Fe = -0.60\,$\mu_{\rm B}$/f.u., Mn = -0.90\,$\mu_{\rm B}$/f.u., V = 0.52\,$\mu_{\rm B}$/f.u. and Al = 0.01\,$\mu_{\rm B}$/f.u., 
leading to a ferrimagnetic structure between $z=0$ and $z=\frac12$ layers along the \textit{c}-direction. However, it may also noted that the theoretical estimate of the magnetic moment does not always quantitatively correspond to the experimentally realized value, but should rather be treated primarily as a qualitative description. Furthermore, vanadium has no independent magnetic moment and the estimated magnetic moment of 0.52 $\mu_{\rm B}$/f.u. at the V-atom can at best be considered as a result of magnetic induction by ordered Mn-spin. In the paramagnetic region where the Mn moments are not yet ordered, vanadium does not exhibit any independent magnetic characteristics and, therefore, the magnetic susceptibility is expected to exhibit a simple Curie-Weiss behaviour controlled primarily by the rather localized magnetic moments of Mn (discussed in Sec.\ref{sec:Magnetism}). The finite line-width of $^{51}${\rm V} NMR spectra  below the Curie temperature confirmed the induced character of magnetic moment at the V-site (discussed in Sec.\ref{sec:NMR}).

It can be seen that DOS exhibits a band-gap at Fermi level (\textit{$E_{\rm F}$}) for the minority (spin-up) band while the majority (spin-down) band is typical of a metal. The band structure for the minority (spin-up) band shows an indirect band-gap. The present calculations show a very high polarization
$P=\frac{\rm{DOS}^\uparrow (E_{\rm F})- \rm{DOS}^\downarrow (E_{\rm F})}{\rm{DOS}^\uparrow (E_{\rm F})+ \rm{DOS}^\downarrow (E_{\rm F})}$ = 92.4\%, indicating that FeMnVAl in the Type~2 ordered structure is nearly a half-metallic ferromagnet.

\subsection{\label{sec:Structure1}Structural analysis - Hypothetical ordered structure}
\begin{figure}
\begin{minipage}{0.49\textwidth}
\includegraphics[width=0.98\textwidth]{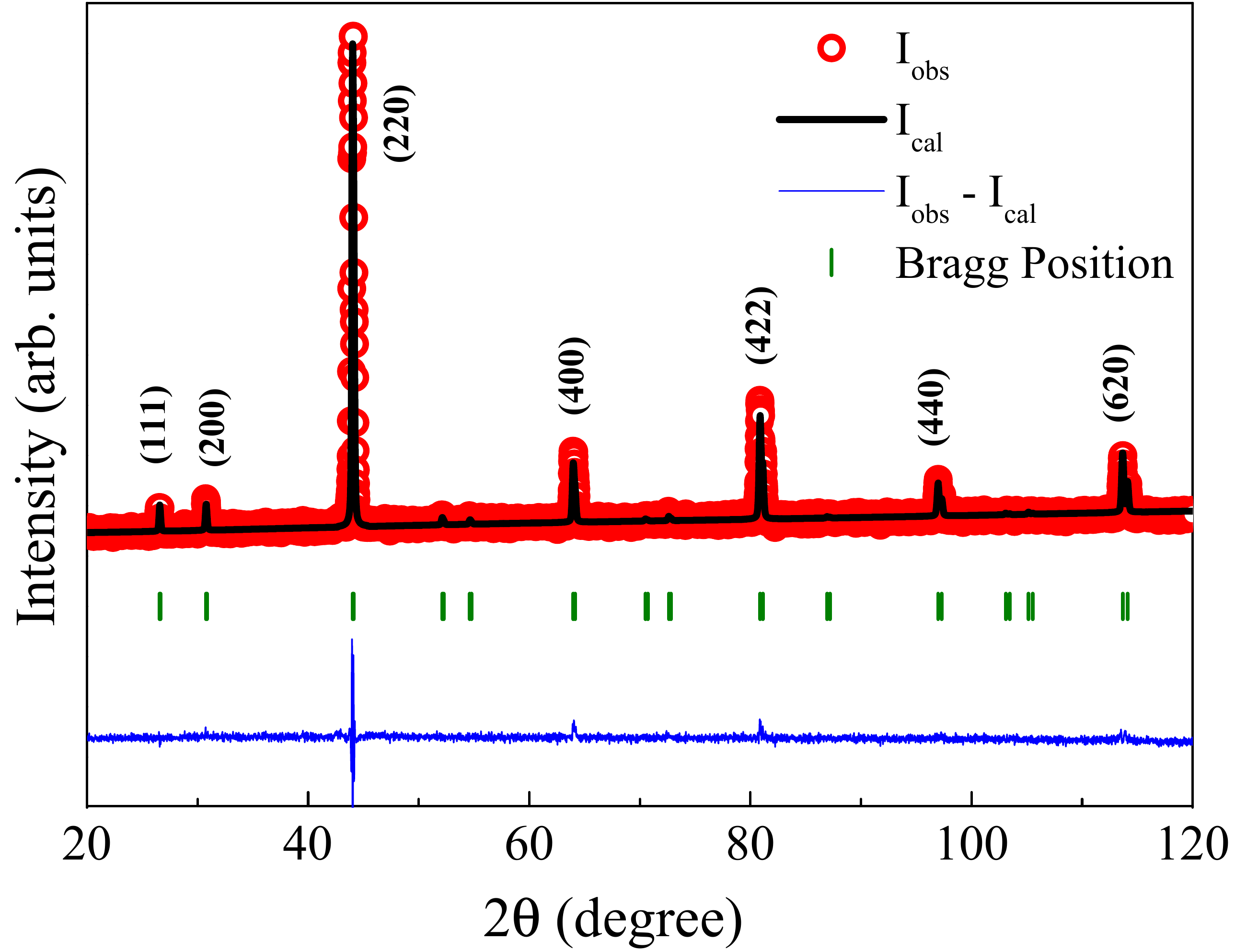}
\caption{Rietveld refinement of the powder XRD pattern measured at room temperature considering Type 2 ordered structure.}
\label{Rietveld}
\end{minipage}
\begin{minipage}{0.49\textwidth}
\includegraphics[width=0.98\textwidth]{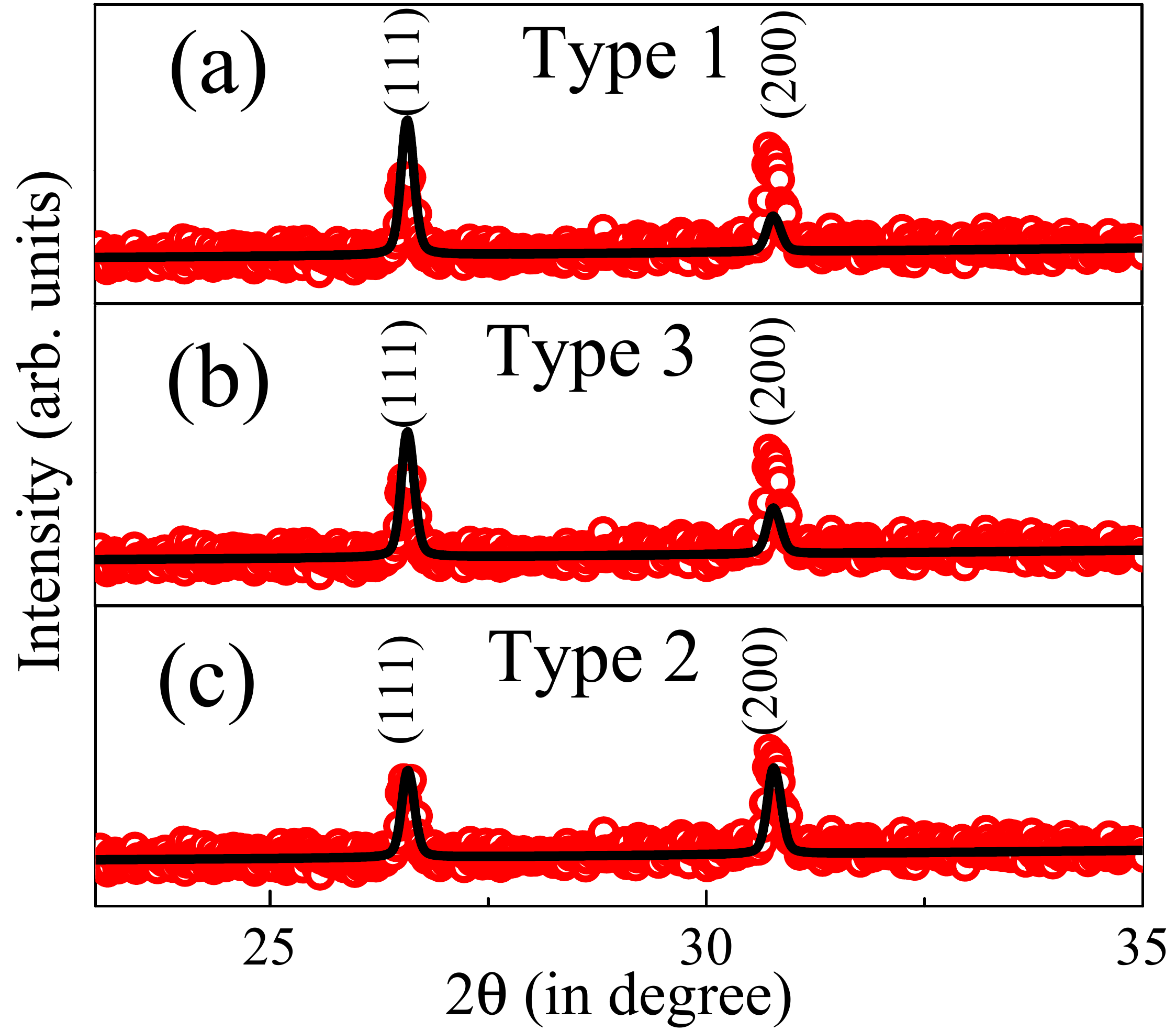}
\caption{Reitveld refinement of the powder XRD assuming (a) Type 1 (b) Type 3 (c) Type 2 structure.}
\label{Different_Structure_XRD}
\end{minipage}
\end{figure}

\begin{figure}[h]
\begin{minipage}{0.49\textwidth}
\includegraphics[width=0.98\textwidth]{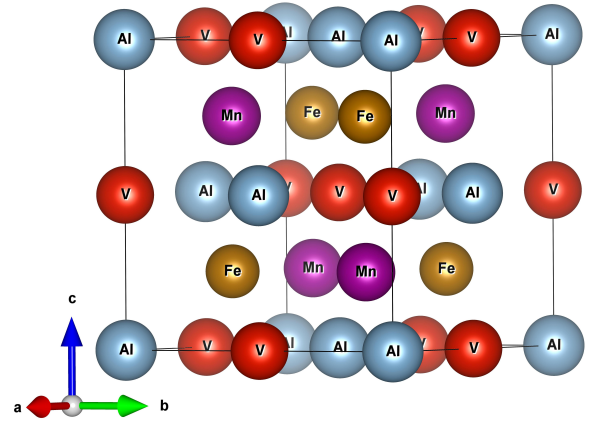}
\caption{Crystal structure of FeMnVAl in Type 2 ordered structure.}
\label{Ordered_Structure}
\end{minipage}
\end{figure}

Fig.~\ref{Rietveld} shows the room temperature Rietveld refinement of the powder XRD pattern. The refinement confirms that the compound crystallizes in a LiMgPdSn-type crystal structure with space group $F\bar{4}3m$ (No. 216). The cubic lattice parameter is $a = 5.821$ {\AA}. The Rietveld refinement further reveals that the experimental data could not be well described by Type 1 and Type 3 structure either, as elaborated in Fig.~\ref{Different_Structure_XRD}. The best fit is obtained with Type 2 structure (represented in Fig.~\ref{Ordered_Structure}), which corresponds to the most stable ordered hypothetical structure calculated by DFT.

A highly ordered structure is one of the main requirements for achieving high spin-polarization in Heusler alloys, while the presence of significant disorder can hinder spin-polarization~\cite{mukadam2016quantification}. A Heusler alloy is generally considered to be fully-ordered when  (111) and (200) superlattice peaks are present in the diffraction pattern. In the presence of disorder, these alloys generally form in disordered structures of the \textit{A2} and \textit{B2}-type ("Strukturbericht"). In case of an \textit{A2}-type structure, both superlattice reflections are absent, while for the \textit{B2}-type disorder, only the (200) peak is present. The presence of peaks (111) and (200) in the XRD pattern of the studied FeMnVAl compound clearly indicates that the crystal structure of the material is in the ordered limit. A rough estimation of the chemical disorder can be made by calculating the ratio  of the peak intensity $I_{(111)}$/$I_{(220)}$ and $I_{(200)}$/$I_{(220)}$~\cite{bainsla2016equiatomic, webster1973magnetic}.

The (111) and (200) reflections of the superlattice are directly proportional to the order parameter, $S^{2}$ and  $S^{2}$$(1-2\alpha)^{2}$, where ~{$S^2$ = [$I_{(200)}$/$I_{(220)}]$$_{\rm(exp.)}$/[$I_{(200)}$/$I_{(220)}]$$_{\rm (theo.)}$} and ~{$S^{2}$$(1-2\alpha)^{2}$ = [$I_{(111)}$/$I_{(220)}$]$_{\rm (exp.)}$/[$I_{(111)}$/$I_{(220)}$]$_{\rm (theo.)}$}. For a well-ordered structure, $S = 1$ and $\alpha = 0$ , and for disordered \rm{A2}- and \rm{B2}-type structures, $S = 0$, $\alpha  = 0$ , and $S = 1$, $\alpha  =0.5$, respectively. The obtained values $S$ and $\alpha$ are found to be 0.980 and 0.0167 for FeMnVAl, which are close to those of the ordered structure. The slight discrepancy, however, may be due to the presence of finite disorder present in the system, which remains below the resolution limit of our XRD studies. In order to detect and confirm the presence of possible disorder, $^{57}$Fe M\"{o}ssbauer measurements were performed in FeMnVAl at 300 K and 77 K (Sec. ~\ref{sec:Mossbauer}).

\subsection{\label{sec:Magnetism}Magnetic properties}
\begin{figure}[h]
\centerline{\includegraphics[width=.50\textwidth]{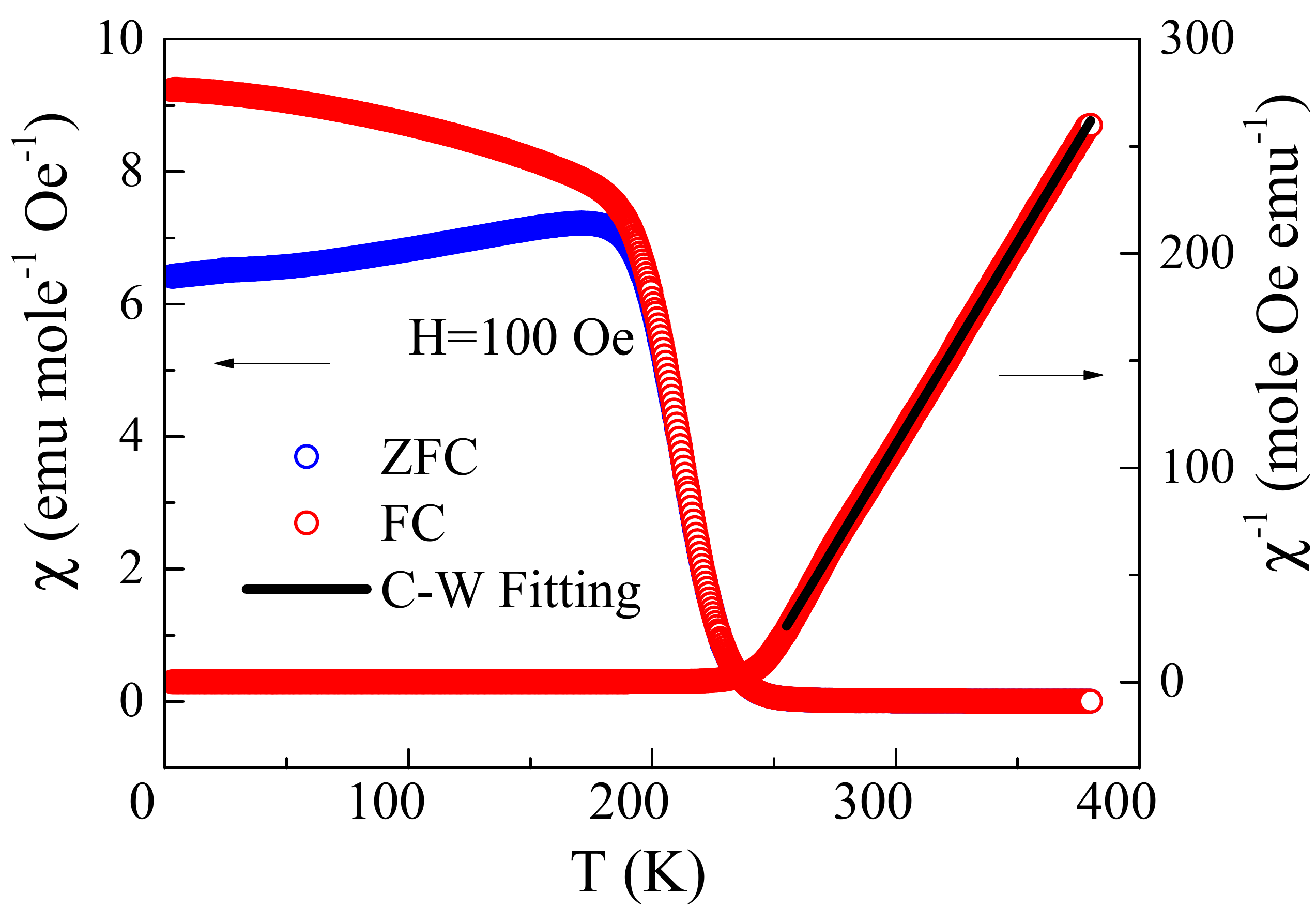}}
{\caption{(Left panel) Temperature dependence of magnetic susceptibility of FeMnVAl measured in a 100 Oe applied magnetic field under ZFC and FC condition. (Right panel) Inverse susceptibility data measured under FC condition. }\label{MT}}.
\end{figure}

The temperature dependence of magnetic susceptibility of FeMnVAl measured in an applied field of 100\,Oe under both zero-field-cooled (ZFC) and field-cooled (FC) conditions, is shown in Fig.~\ref{MT}. The compound undergoes a paramagnetic (PM) to ferromagnetic (FM) phase transition below $T_{\rm C} = 213$~K ($\pm{0.5}$)\@. The $T_{\rm C}$ is determined as the temperature at which $d\chi/dT$ exhibits a minimum (data not shown here). Thermomagnetic irreversibility is observed between ZFC and FC susceptibilities below $T < T_{\rm C}$, indicating the critical field value (H$_c$) of the isothermal hysteresis is greater than 100\,Oe. The thermomagnetic irreversibility vanishes with an application of 500\,Oe (data not shown here). Above the ordering temperature magnetic susceptibility data follows Curie-Weiss (CW) law given by ${\chi= C/(T-\theta_P)}$, where \rm C is the Curie constant and $\theta_P$ is paramagnetic Curie temperature. The linear CW fit to the inverse susceptibility (Fig.~\ref{MT})  in the region 250-380 K yields $\theta_P$ = +241.3(3) K. The positive sign of $\theta_P$ further confirms the presence of ferromagnetism in this material. In order to verify whether the magnetism of this compound is localized (dominated by Mn-spin) or itinerant-type, we determined the Rhodes-Wohlfarth ratio (RWR)~\cite{saunders2020exceedingly}. RWR is defined as the ratio of P$_C$/P$_S$, where P$_C$ is the paramagnetic moment obtained as $\mu_{effective}^2$ = P$_C$(P$_C$+2) and P$_S$ is the saturation magnetization at low temperature. For the localized moment system, RWR is close to 1 and for conventional itinerant system RWR is found to be greater than unity~\cite{saunders2020exceedingly}. For FeMnVAl, RWR is estimated to be 1.51 which is quite low compared to other reported conventional itinerant ferromagnets~\cite{saunders2020exceedingly,bhattacharyya2011investigation,mondal2021critical}, but rather close to unity as expected in local moment systems. To further validate the non-dominant nature of itinerant magnetism, we have also plotted (figures not shown here) M$_{S}^{2}$ \textit{vs.} T$^2$ below Curie temperature ($T_{\rm C}$) and M$^4$ \textit{vs.} H/M near $T_{\rm C}$. According to the self-consistent renormalization (SCR) theory for itinerant electron magnetism, the above two plots should be linear in nature~\cite{moriya1978spin,takahashi1986origin}. The deviation from linearity for both the curves in the present case confirms that the nature of the magnetism in the studied compound is not of itinerant type, but dominated by the localized Mn-spin moment.

\begin{figure}[h]
\centerline{\includegraphics[width=.48\textwidth]{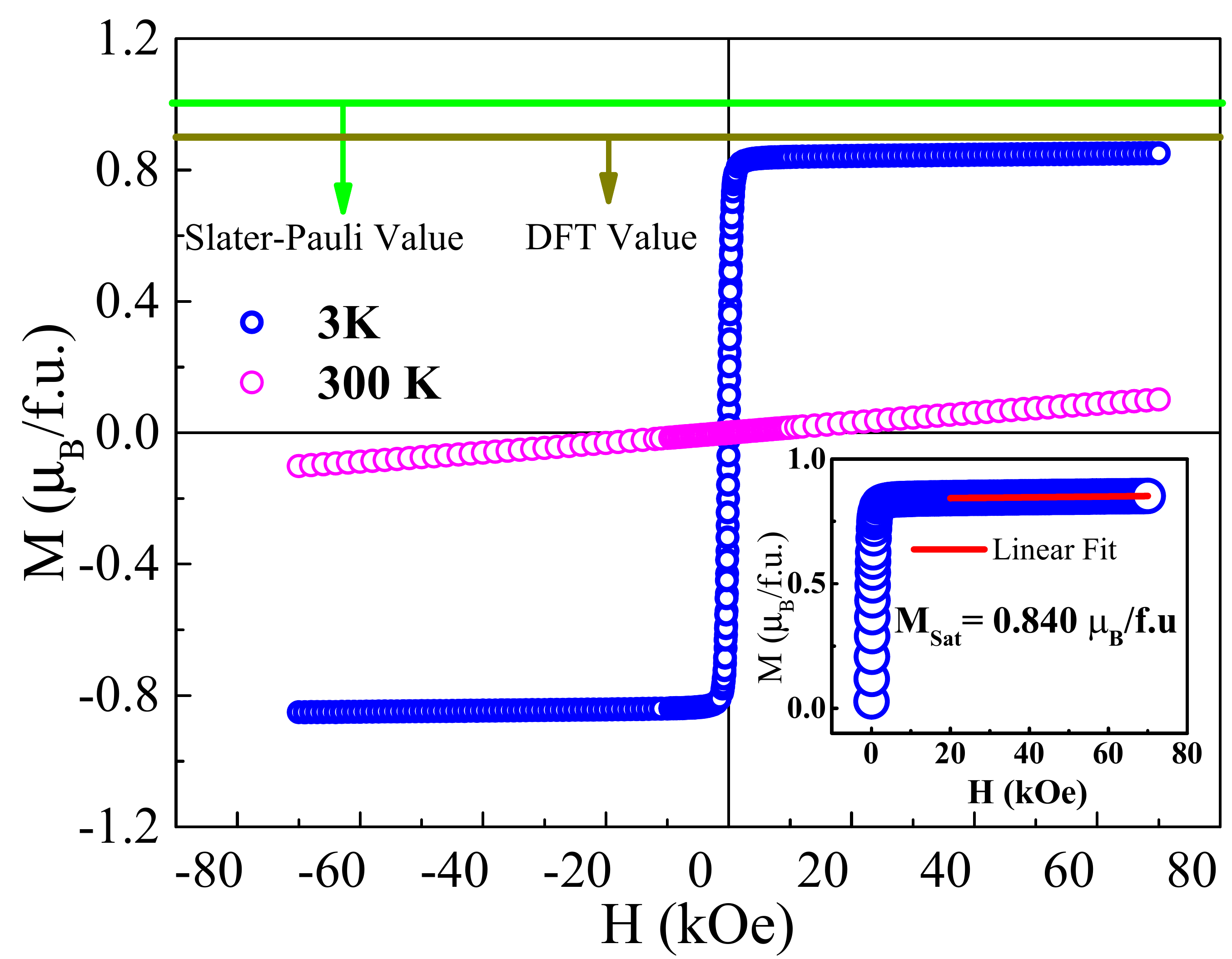}}
{\caption{ Isothermal magnetization of FeMnVAl measured at 3 K and 300 K. Solid line represent the Slater-Pauling value. Inset shows the saturation magnetization fitting.}\label{MH}}
\end{figure}

 As mentioned earlier, the total VEC for FeMnVAl is 23, and therefore the total magnetic moment should be -1 $\mu_{\rm B}$/f.u\@. To verify the applicability of the S-P rule, isothermal magnetization measurements as a function of field were performed for the system. Fig.~\ref{MH} shows the $M(H)$ behavior measured at 3\,K ($T < T_{\rm C}$) and 300\,K ($T > T_{\rm C}$). FeMnVAl shows a soft ferromagnetic-like behavior with negligible hysteresis ($\sim$125 Oe), confirming the thermomagnetic irreversibility of the magnetic susceptibility measured only under a low magnetic field (H$_C$ $> $100 Oe). The saturation moment at 3 K, as estimated by linear extrapolation of the high field magnetization data shown in the inset of Fig.~\ref{MH}, is found to be $M_{\rm sat} = 0.84 \, \mu_{\rm B}$/f.u.\@. The origin of such deviation could be due to the presence of minor disorder in the compound studied,  as suggested previously in the analysis of the XRD pattern. The estimated value of $M_{\rm sat}$ is slightly lower than the expected Slater--Pauling moment value for this compound with $N_V = 23$. This decrease in magnetic moment resulting from structural disorder is also supported by our spin-polarized band structure calculations discussed later in Sec.~\ref{sec:ElectronicPart2}.

\subsection{\label{sec:Mossbauer}M\"{o}ssbauer spectrometry}

\begin{figure}[h]
\centerline{\includegraphics[width=.48\textwidth]{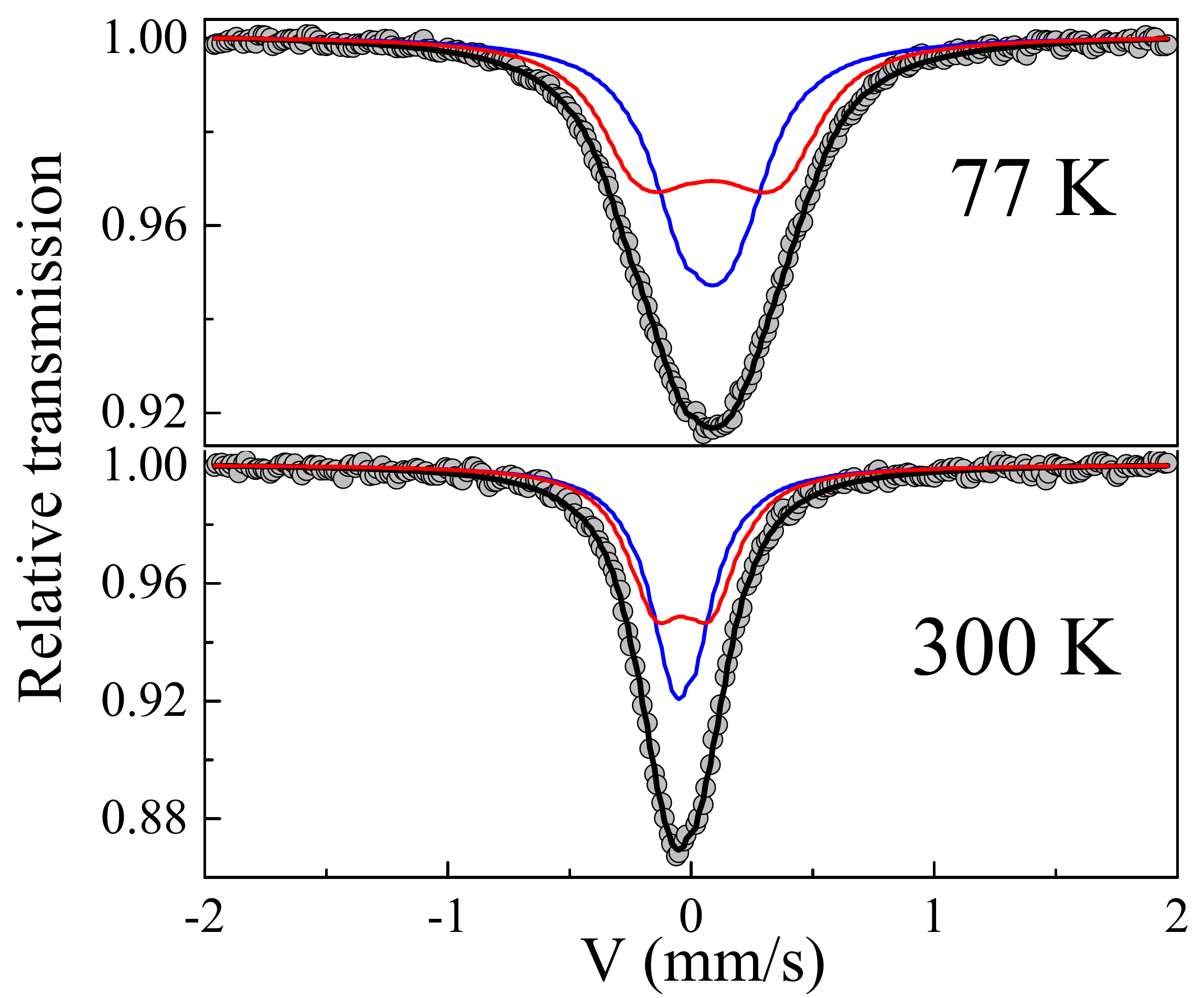}}
{\caption{M\"{o}ssbauer spectra of FeMnVAl taken at 300 K and 77 K.}\label{Fig_Mossbauer}}
\end{figure}
In order to better understand the structural disorder and magnetism in FeMnVAl, $^{57}$Fe M\"{o}ssbauer measurements were performed at 300 K and 77 K. The spectra show a single broadened and asymmetrical line which cannot be well described by a single Lorentzian line. As shown in Fig.~\ref{Fig_Mossbauer}, the  spectrum at 300 K can be better described by considering at least two quadrupolar doublets with small values of quadrupolar splitting which is consistent with the nearly cubic symmetric environment of the Fe atom. At 77 K, the significantly broadened spectrum can also be described by two quadrupolar components but the increase in the quadrupolar strength cannot be reasonably explained.  The fitted parameters of the Mössbauer spectra  are presented in Table~\ref{Mossbauer}. One of the simplest possibilities is to consider that the spectrum to be described by two magnetic components, nearly of equal intensities. The low values of hyperfine fields (0.7 and 1.9 T) indicate that the Fe moments participate in the ferromagnetic ordering but the major contribution to the total magnetic moment comes from the Mn moment, that is also consistent with the theoretical calculations. Although in the ordered structure of Type 2, Fe has a single crystallographic site, the appearance of two quadrupolar doublets clearly indicates the presence of two sites for Fe atom having similar structural environment. In ordered Type-2 structure of FeMnVAl, the Mn (4\textit{c} site) and Fe (4\textit{d} site) atoms have an equivalent environment consisting of 4 V + 4 Al as nearest neighbour (NN). Similarly, Al (4\textit{a} site) and V (4\textit{b} site) atoms also have  identical environments which consist 4 Fe + 4 Mn atoms as NN. For the studied compound, Fe occupies 4\textit{d} position in the Type 2 structure, but the presence of two doublets in M\"{o}ssbauer spectra suggests the presence of Fe at 4\textit{c} position as well, in consonance with the 4\textit{c} and 4\textit{d} positions having a similar environment. The two M\"{o}ssbauer components are a priori rather equiprobable but it is important to note that the lack of resolution of the total hyperfine structure prevents a physically accurate estimation.
\begin{table}[]
\caption{ Fitted parameter values for the M\"{o}ssbauer of FeMnVAl. Isomershift ($\delta$) , Linewidth at half height (${\Gamma}$) (quoted relative to $\alpha$-Fe at 300 K), quadrupolar shift (${\frac {Q} {2\varepsilon}}$), hyperfine field (B$_{hf}$) and relative proportions ($\%$) are estimated at 300 K and 77 K.}
\begin{tabular}{|c|c|c|c|c|c|c|}
\hline
\hline
 T (K) & Site & $\delta$ (mm/s) & ${\Gamma}$ (mm/s) & ${\frac {Q} {2\varepsilon}}$  & B$_{hf} (T) $  & $\%$     \\
 & & $\pm$0.01 &$\pm$0.01 &$\pm$0.01 &$\pm$0.3 &$\pm$2\\ \hline
300    & Fe1 & 0.08 & 0.28 & 0.01  & - & 50                         \\
           & Fe2 & 0.09 & 0.28 & 0.21  & - & 50                         \\ \hline
77 K  & Fe1 & 0.20 & 0.36 & 0.00  &0.7 & 48                         \\
         & Fe2 & 0.20 & 0.36 & 0.00  & 1.9& 52                         \\
\hline
\end{tabular}
\label{Mossbauer}
\end{table}

\subsection{\label{sec:Structure2}Structural analysis - Disordered structure}

\begin{figure}
\begin{minipage}{0.49\textwidth}
\includegraphics[width=0.98\textwidth]{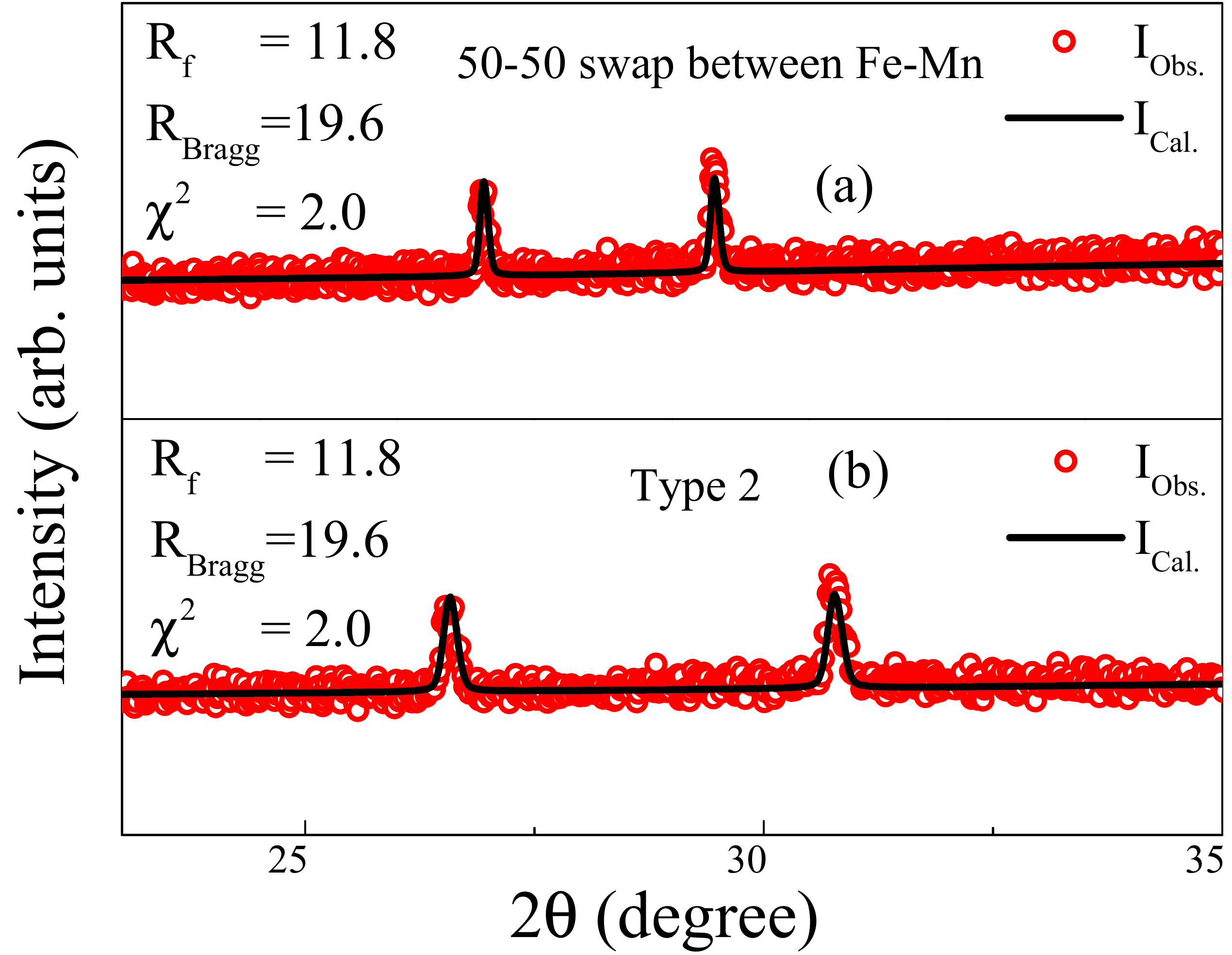}
\caption{Reitveld refinement of the powder XRD assuming (a) disordered (50-50 swap between Fe-Mn; (L2$_1$-type crystal structure) (b) Type 2 ordered structure (Y-type crystal structure).}
\label{Disorder_XRD}
\end{minipage}
\begin{minipage}{0.49\textwidth}
\includegraphics[width=0.98\textwidth]{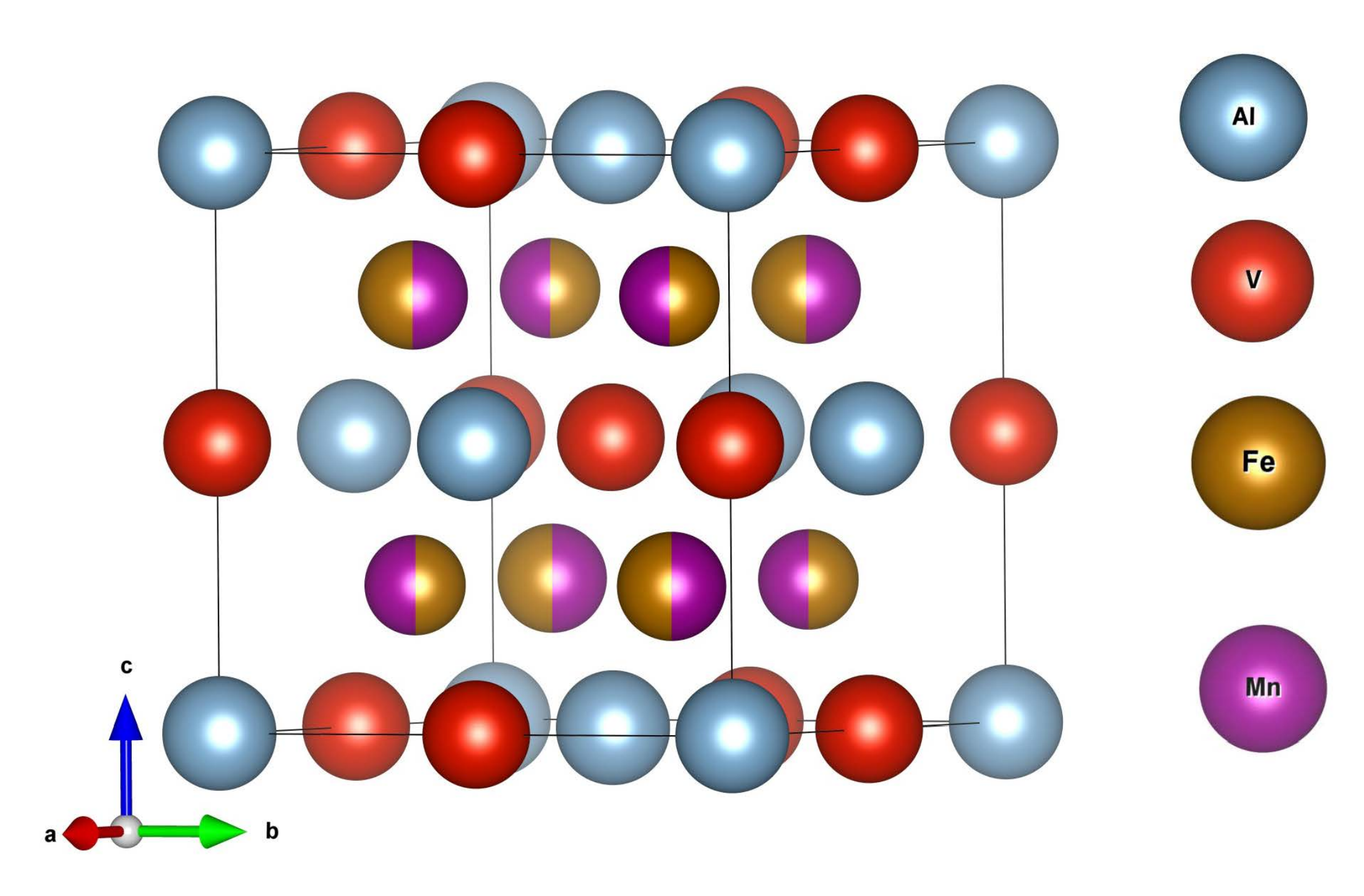}
\caption{Crystal structure \textcolor{blue}{(L2$_1$-type)} of FeMnVAl in disordered structure.}
\label{Disorder_Structure}
\end{minipage}
\end{figure}
As the analysis of M\"{o}ssbauer spectra asserts the presence of two different Fe-sites with 50\% occupancy at each sites, we must revisit the XRD analysis presented earlier Sec.~\ref{sec:Structure1} where we had considered a completely ordered crystal structure with only one site occupancy for Fe. In the ordered structure (Type 2) Fe occupies 4\textit{d} position, while the M\"{o}ssbauer spectra indicate presence of two sites for Fe atoms having 50/50 occupancy in each site. Subsequently, we checked the possibility of the presence of the Fe atom at  two sites by exchanging Fe with other atoms in all possible combinations, namely (4$a$,4$d$), (4$b$,4$d$) and (4$c$,4$d$), respectively. The Rietveld refinements were performed by considering all the possible combinations, but the quality of the fit deteriorates for both the 4$a$ \& 4$d$  position permutation types as well as 4$b$ \& 4$d$ for 50\% cross-swapping of the respective occupants. On the other hand, the quality of fit remains essentially invariant to the ordered structure when 50\% of Fe (4\textit{d}) is placed in Mn (4$c$)-site and vice-versa (Fig.~\ref{Disorder_XRD}). As Mn and Fe are neighbouring elements in the periodic table, they have very close X-ray scattering cross-sections and hence the XRD analysis therefore can not distinguish the two different configurations. The XRD analysis, however, establishes that if the Fe atoms are distributed in two different sites, it can only be between the 4\textit{c} and 4\textit{d} sites. Interestingly, due to the 50:50 exchange between the Fe (4\textit{d} sites) and Mn (4\textit{c} sites) atoms in the Y-type crystal structure (space group: $F\bar{4}3m$, no. 216 ), the resultant structure reverts to the more symmetric L2$_1$-type crystal structure (space group: $Fm\bar{3}m$, no. 225) describing the ordered structure for full Heusler alloy. We would like to point out here that in a later section (~\ref{sec:ElectronicPart2}), it was established that the formation energy of the disordered structure (represented in Fig~\ref{Disorder_Structure}) is lower in comparison to that of the ordered structure (Type-2). The two Fe sites (4$c$ and 4$d$) in this disordered structure having a similar environment, which is consistent with the results obtained by M\"{o}ssbauer spectrometry.

\subsection{\label{sec:NMR}Nuclear Magnetic Resonance (NMR)}
\begin{figure}[h]
\centerline{\includegraphics[width=.48\textwidth]{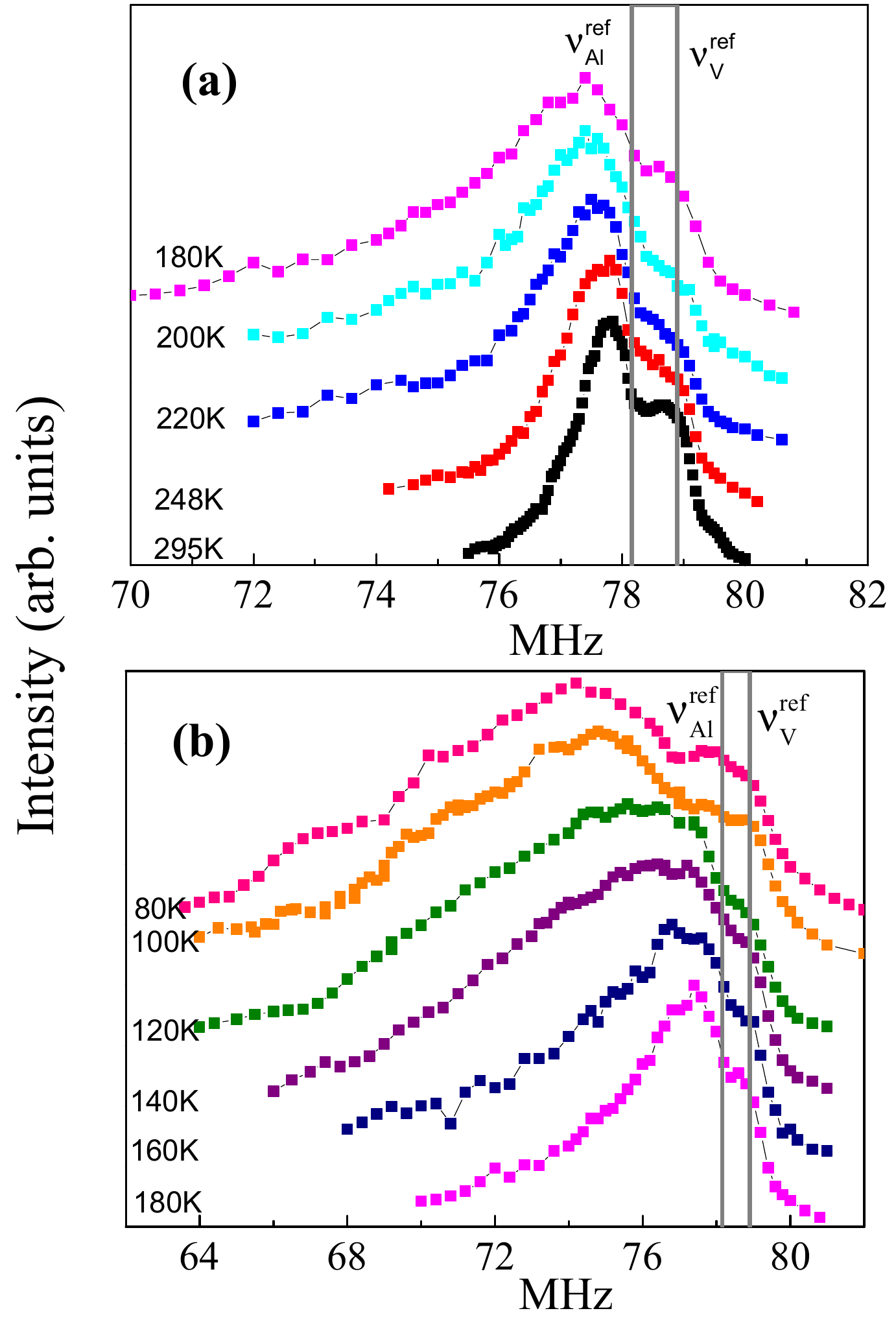}}
{\caption{NMR spectra of FeMnVAl at various temperatures. The vertical lines indicate the reference frequency positions of $^{27}$Al and $^{51}$V.}\label{NMR1}}
\end{figure}

 Fig.~\ref{NMR1} shows the NMR spectra of FeMnVAl at various temperatures 80-295 K with respect to the position of reference frequencies of the ligand nuclei $^{27}${\rm Al}  ($\nu^{ref}_{\rm Al}$ = 78.157 MHz) and $^{51}${\rm V}  ($\nu^{ref}_{\rm V}$ = 78.895 MHz). Two distinct peaks are observed in the spectra taken at room temperature (295 K). As the temperature is lowered, a considerable broadening of the spectra is observed, accompanied by an overall shift towards the lower frequencies. It is interesting to note that at 80 K the spectrum is distributed over a frequency range of about 18 MHz, which is much larger than the spectral distribution of about 1 MHz in ternary Heusler alloy systems~\cite{ooiwa1998nuclear, suh2006antisite}. Such a large temperature-dependent broadening and shift of the resonance line indicates a hyperfine electron-nuclear interaction and a magnetic dipolar interaction with intrinsic localized electronic magnetic moments in the system.

A close examination of the spectra reveals that the part of spectra near the reference positions, $\it{i.e.}$, around 77-82 MHz is rather weakly affected by the temperature variation compared to the low-frequency part where the broadening and the shift are much more pronounced. In other words, the spectra at all temperatures may have a narrow component which is largely unaffected by temperature variation, and a broad component for which the position and the width depend upon temperature. This clearly indicates that part of the sample is non-magnetic whereas the other part is magnetic, pointing towards the inhomogeneous magnetic nature of the sample.

Vanadium atoms are non-magnetic in these Heusler alloy systems as already known from an earlier study of  $^{51}${\rm V} NMR in Fe$_2$VSi~\cite{nishihara2004nmr} and also from the study of both $^{27}${\rm Al} and $^{51}${\rm V} NMR in Fe$_2$VAl~\cite{suh2006antisite}. In the latter study, $^{27}${\rm Al} and $^{51}${\rm V} resonance lines were narrow and well-resolved, and so it was possible to carry out spin-lattice relaxation time measurements for both $^{27}${\rm Al} and $^{51}${\rm V} resonance. The results indicated that even though $\nu^{ref}_{\rm Al} < \nu^{ref}_{\rm V}$, the resonance of $^{27}$Al occurred at a higher frequency than that of $^{51}$V. It was thus established that $^{27}${\rm Al} has a positive shift whereas $^{51}${\rm V} resonance is affected by a stronger and negative shift in Fe$_2$VAl. In metallic alloys, the dominant contribution in the shift of nuclear resonance position is produced by the Fermi contact interaction with the conduction electrons, and is called the Knight shift. In presence of localized magnetic moments, the polarization of conduction electrons may produce a negative Knight shift in ligand atomic nuclei. However, owing to the presence of $d$-electrons and consequent strong core electron polarization of vanadium atoms, $^{51}${\rm V} nuclei should experience a larger negative magnetic hyperfine field in comparison to the small field in $^{27}${\rm Al}~\cite{suh2006antisite}.

In the present case, it is therefore reasonably assumed that the NMR spectra arise from both $^{27}${\rm Al} and $^{51}${\rm V} resonances, with a narrow and a broad component for both the resonances. So, the spectra at all temperatures have been simulated as a sum of four components. The various resonance frequencies ($\nu$) have been taken as,

\begin{eqnarray}
\nu^{dia}_{\rm Al} = \nu^{ref}_{\rm Al} (1+K^{dia}_{\rm Al}) \nonumber  \\
\nu^{dia}_{\rm V} = \nu^{ref}_{\rm V} (1+K^{dia}_{\rm V}) \nonumber  \\
\nu^{mag}_{\rm Al} = \nu^{ref}_{\rm Al} (1+K^{mag}_{\rm Al}) \nonumber  \\
\nu^{mag}_{\rm V} = \nu^{ref}_{\rm V} (1+K^{mag}_{\rm V})
\label{eqn2}
\end{eqnarray}
\noindent
where, $\it{dia}$ denotes the resonance corresponding to the weakly magnetic or diamagnetic environment and $\it{mag}$ is that to the magnetic environment. It follows from the considerations described above that the various $K$, the isotropic shift of the respective components, should be related to each other as,

\begin{equation}
K^{\textit{dia}}_{\textit{\rm Al}}>0,\ \rm{and,  }K^{\textit{mag}}_{\rm V}<K^{\textit{mag}}_{\textit{\rm Al}}<K^{\textit{dia}}_{\textit{\rm V}}<0
\end{equation}

In deconvolution, we have used line broadening $f(\nu)\propto\exp(\frac {-(\nu-\nu_i)^2}{(\omega^G_i)^2})$, for the non-magnetic components, and Lorentzian broadening, $f(\nu) \propto\frac{\omega^L_i}{1+[(\nu-\nu_i)\omega^L_i]^2}$ for the paramagnetic components. In these expressions, $\nu_i$ are the resonance frequencies as in the left hand side of eqn.~\ref{eqn2}, $\omega^{G}_{i}$ is the corresponding linewidth parameter when the resonance component is Gaussian, and $\omega^{L}_{i}$ is the linewidth parameter when the resonance component is Lorentzian.

\begin{figure}[h]
\centerline{\includegraphics[width=.48\textwidth]{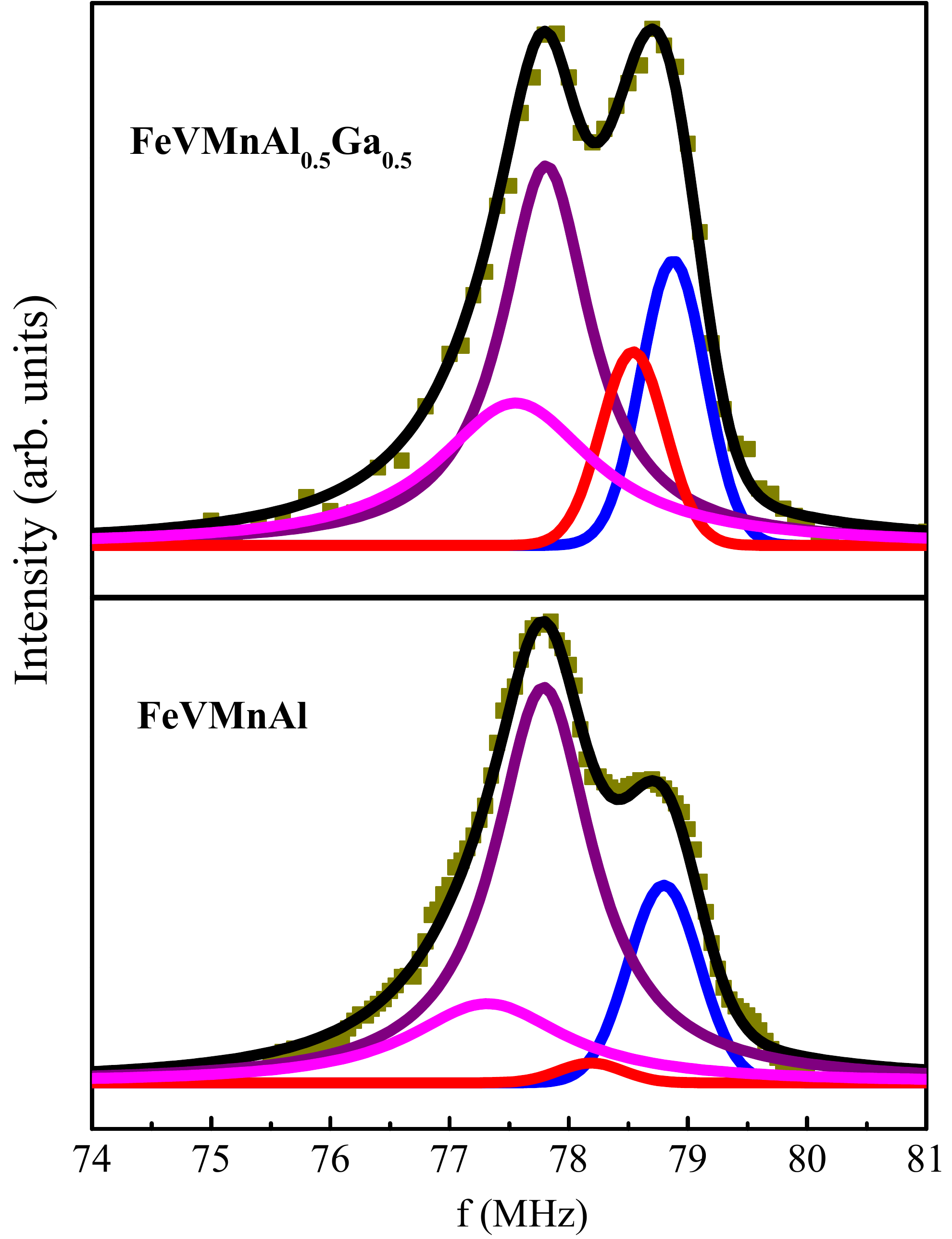}}
{\caption{NMR spectra (solid squares) at 295 K of FeVMnAl and FeVMnAl$_{0.5}$Ga$_{0.5}$ , and their simulations as sum of  component lines corresponding to diamagnetic and magnetic environments, shown as, {\rm Al}$_{dia}$ (blue), {\rm Al}$_{mag}$ (purple), {\rm V}$_{dia}$ (red), {\rm V}$_{mag}$ (magenta).}\label{NMR2}}
\end{figure}

Fig.~\ref{NMR2} shows the result of the simulation of the spectrum at 295 K. However, in order to further confirm the positions of $^{27}${\rm Al} and $^{51}${\rm V} resonances, we prepared a sample with partial substitution of Al with Ga, $\it{i.e.}$, FeMnVAl$_{0.5}$Ga$_{0.5}$. Here, we present the NMR spectra of FeMnVAl$_{0.5}$Ga$_{0.5}$ that has shown magnetization behavior and overall room temperature NMR linewidth  similar to those of FeMnVAl. The NMR spectra of both these samples are simulated and shown in Fig.~\ref{NMR2}. $^{51}${\rm V} resonance components, as designated, become stronger compared to $^{27}${\rm Al}  resonances when Al is partially replaced by Ga, thus confirming the positions of $^{27}${\rm Al} and $^{51}${\rm V} resonances in the composite spectrum of FeMnVAl.

\begin{figure}[h]
\centerline{\includegraphics[width=.48\textwidth]{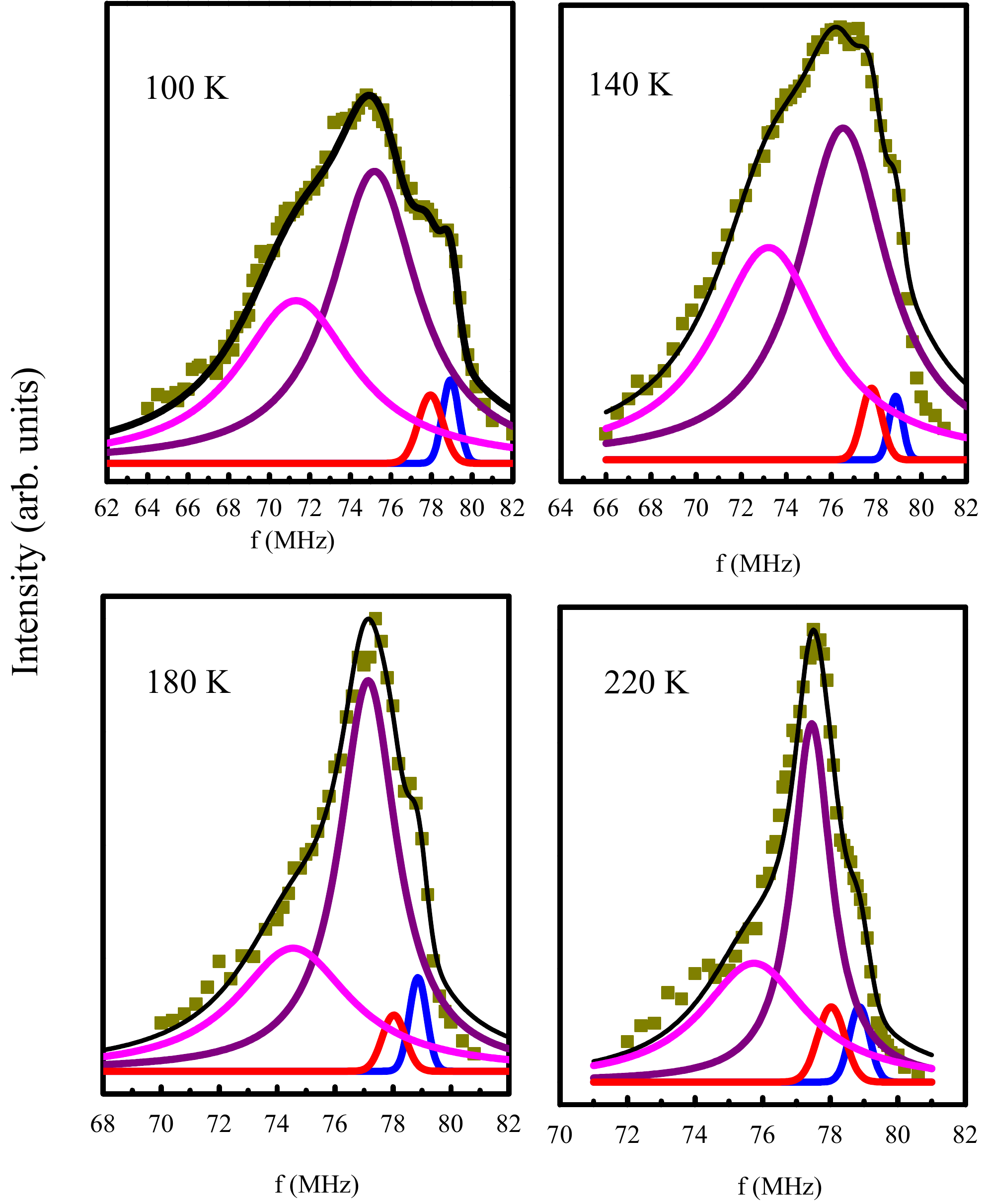}}
{\caption{NMR spectra (solid squares) at various temperatures of FeVMnAl, and their simulations as sum of component lines corresponding to diamagnetic and magnetic environments, shown as, {\rm Al}$_{dia}$ (blue), {\rm Al}$_{mag}$ (purple), {\rm V}$_{dia}$ (red), {\rm V}$_{mag}$ (magenta).}\label{NMR3}}
\end{figure}

The simulation of the spectra at various temperatures 80-295 K are shown in Fig.~\ref{NMR3}. Even at lower temperatures where the spectra are quite broad, the simulated spectra well fit the experimental ones. It should be mentioned that anisotropic magnetic broadening was not considered in the simulation. Moreover, it is assumed that structural disorder as obtained here does not significantly alter the characteristic local cubic symmetry of Heusler alloy systems. Therefore, the broadening effect of nuclear quadrupolar interaction was neglected.

\begin{figure}[h]
\centerline{\includegraphics[width=.48\textwidth]{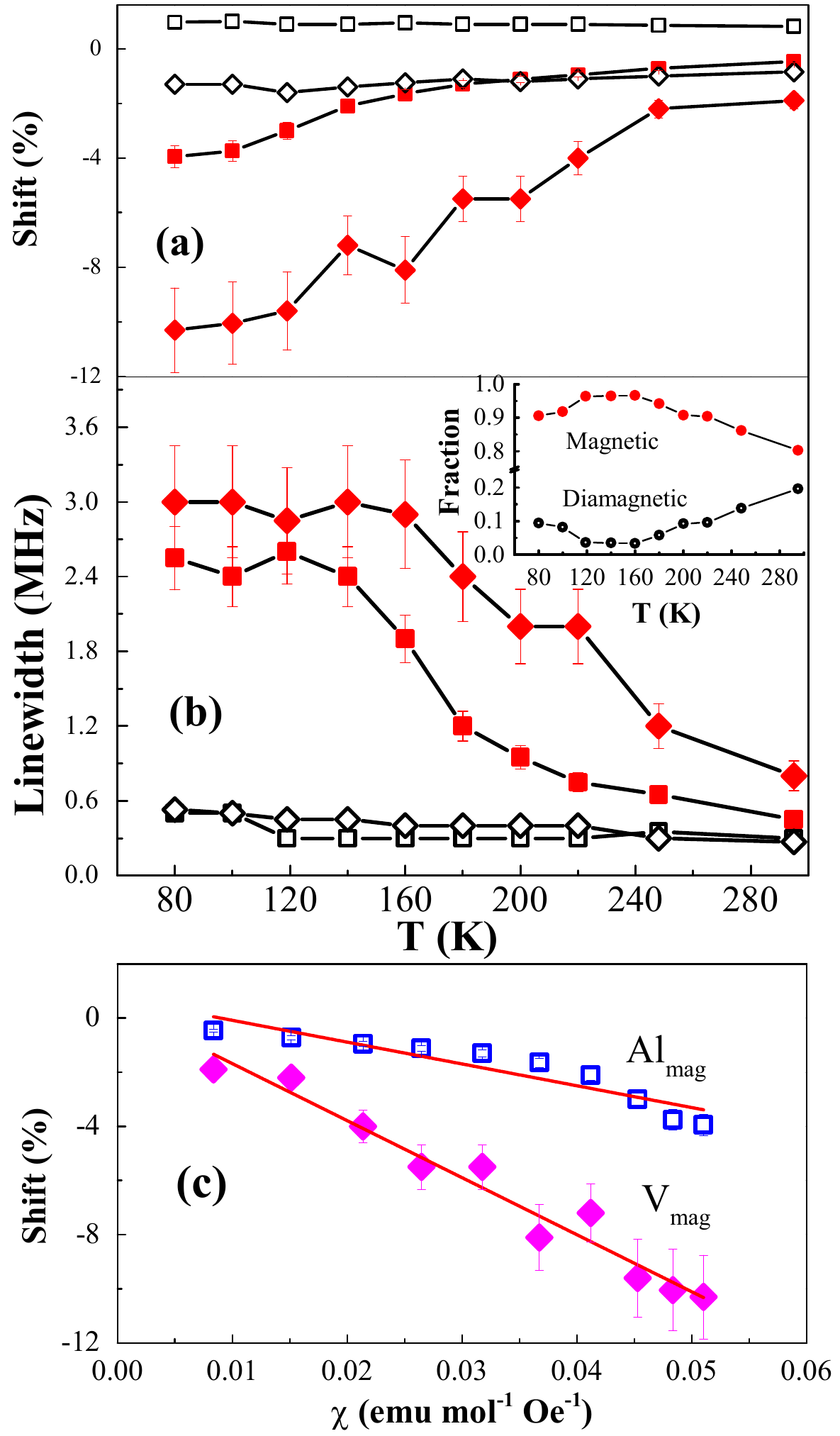}}
{\caption{Results of analysis of NMR spectra. The parameters corresponding  to different components in the spectra are denoted as, {\rm Al}$_{dia}$ (square), {\rm V}$_{dia}$ (diamond), {\rm Al}$_{mag}$ (solid square), {\rm V}$_{mag}$ (solid diamond). Error bars are shown when they are significantly large. (a) Temperature dependence of shift. (b) Temperature dependence of linewidth and inset represents the contributions of diamagnetic and magnetic fractions. (c) Shift vs. bulk magnetic susceptibility $\chi^{exp}$ of FeVMnAl. In (a) and (b), the lines are guide to the eye. In (c), the lines denote fit to the equation 4.}\label{NMR4}}
\end{figure}

The temperature dependence of the line widths of the four resonance components and their positions are presented in Fig.~\ref{NMR4}. The positions and linewidths of the diamagnetic resonance components show only a small variation, while those of the magnetic components show a large variation over the temperature range. The resonance line width is determined by the temperature-dependent time-averaged fluctuation of the local magnetic field. As expected, the line widths of the diamagnetic resonance components  $^{27}$Al and  $^{51}$V are not significantly influenced by the magnetism of the system. On the other hand, the linewidths of the magnetic components increase with decreasing temperature, thus mimicking the magnetization.

Over the entire temperature range $K^{dia}_{\rm Al}$ is positive and shows almost no variation, with values between 0.9 to 1.0 \%. $K^{dia}_{\rm V}$ is negative and shows only a small decrease with decreasing temperature, varying slowly from -0.4 to -1.7 \%. However, for magnetic components, both $K^{mag}_{\rm Al}$ and $K^{mag}_{\rm V}$ show a pronounced decrease with decreasing temperature.

In metallic systems where non-magnetic ligand nuclei experience a hyperfine field that is influenced by localized magnetic moments, $K$ is expressed as a sum of (a) a temperature independent part ($K^0$) that is proportional to the susceptibilities of core electron diamagnetism, orbital magnetism and pauli paramagnetism; and, (b) a temperature dependent part ($K(T)$) that is proportional to magnetic susceptibility ($\chi^{spin}$) due to the localized spin magnetic moments, as given by,

\begin{equation}
K =K^0+( {H^{hf}} /{N_A\mu_B})\chi^{spin}(T)
\end{equation}
\noindent
where $H^{hf}$ is the hyperfine field at the nucleus. In the presence of localized moments, $\chi^{spin}$ is almost identical to the experimental bulk magnetic susceptibility ($\chi^{exp}$). We have taken $\chi^{exp}$ values measured at the magnetic field of 70 kOe, $\it{i.e.}$, about the same field as used in NMR measurements. The linear fit of $K$ $vs.$ $\chi^{exp}$ data as shown in Fig.~\ref{NMR4}(c) yields, $K_{\rm Al}^{0}$$\simeq{0.007}$($\pm{0.002}$)\%
and $H^{hf}$ of -{4.5}($\pm{0.3}$) kOe; and, $K_{V}^{0}$$\simeq{0.004}$($\pm{0.003}$)\% and  $H^{hf}_{\rm V}$ of -{11.8}($\pm{0.6}$) kOe. These values of $H^{hf}$ are comparable in order of magnitude to the hyperfine fields obtained for $^{27}$Al in ferromagnetic Co$_2$TiAl~\cite{shinogi1984hyperfine,grover1979paramagnetic}, and also for $^{27}$Al and $^{51}$V in ferromagnetic Co$_2$VAl~\cite{yoshimura1985hyperfine}. In the present study, the linearity of $K$ $vs.$ $\chi$ is maintained throughout the temperature range, $\it{i.e.}$, the hyperfine field remains the same above and below the Curie temperature. It may be noted here that in some other magnetic Heusler alloys too, \textit{e.g}. Co$_2$VAl~\cite{yoshimura1985hyperfine}, the transferred hyperfine fields remain almost same in both ferromagnetic as well as paramagnetic states.

The deconvolution procedure allows us to roughly estimate the contributions of different types of probe nuclei in the composite spectrum. The temperature variation of the integrated intensities of the non-magnetic (combined for $^{\rm 27}$Al and $^{\rm 51}$V) and magnetic (combined for $^{\rm 27}$Al and $^{\rm 51}$V) components (see Fig.~\ref{NMR4} for details) is shown in the inset of Fig.~\ref{NMR4}(b). It shows that from about 200 K down to the lowest observed temperature, about $\sim 10\%$ of the probe nuclei undergo almost diamagnetic local environment though it is an overwhelmingly magnetic system. To understand the presence of such  magnetic and non-magnetic components, we use the following model.

M\"{o}ssbauer spectra and XRD results confirmed 50/50 exchange in the 4$c$ (0.25,0.25,0.25) and 4$d$ (0.75,0.75,0.75) position between Fe and Mn atoms. The M\"{o}ssbauer spectra also confirm that Fe atoms show no long range magnetic order, even below T$_C$, and the main  contribution of the total magnetic moment comes mainly from the Mn site, which is also verified by the band structure calculations. The crystal structure of FeMnVAl suggests that Al and V occupy octahedral positions and that both have similar environments due to the symmetry of the crystal structure. As there is  a random distribution of Fe and Mn atoms in the 4$c$ and 4$d$ sites, the atomic surroundings for Al and V are subject to local environmental variation. From the macroscopic perspective, the nearest neighbors (NN) of Al (V) are 4 Mn and 4 Fe for a 50/50 exchange between Fe and Mn at a distance ${\sqrt3a}$/4. However, as the Mn/Fe arrangement in this disordered structure is random, the Al(V) sites will experience a variety of different local environments. In the present case, the Al(V) sites could have 9 different nearest-neighbour local environments: 8 Fe+0 Mn, 7 Fe+1 Mn, 6 Fe +2 Mn, 5 Fe+3 Mn, 3 Fe+5 Mn, 2 Fe+6 Mn, 1 Fe+7 Mn and 0 Fe+8 Mn. Since Mn is the major contributor of the total magnetism and Fe atoms do not carry magnetic moment in the studied compound, the Al(V) environment dominated by the Fe atoms (8 Fe+0 Mn, 7 Fe+1 Mn, 6 Fe+2 Mn, \textit{etc.}) likely to remain nonmagnetic and responsible for the observed 10 \% nonmagnetic component in the NMR spectra. A similar variation of the local environment was earlier observed in the NMR spectra of Co$_2$Mn$_{1-x}$Fe$_x$Si~\cite{wurmehl2007probing,wurmehl201355}.

\subsection{\label{sec:ElectronicPart2}Electronic structure calculations- Disordered structure}
\begin{figure*}[ht]
\centerline{\includegraphics[width=.96\textwidth]{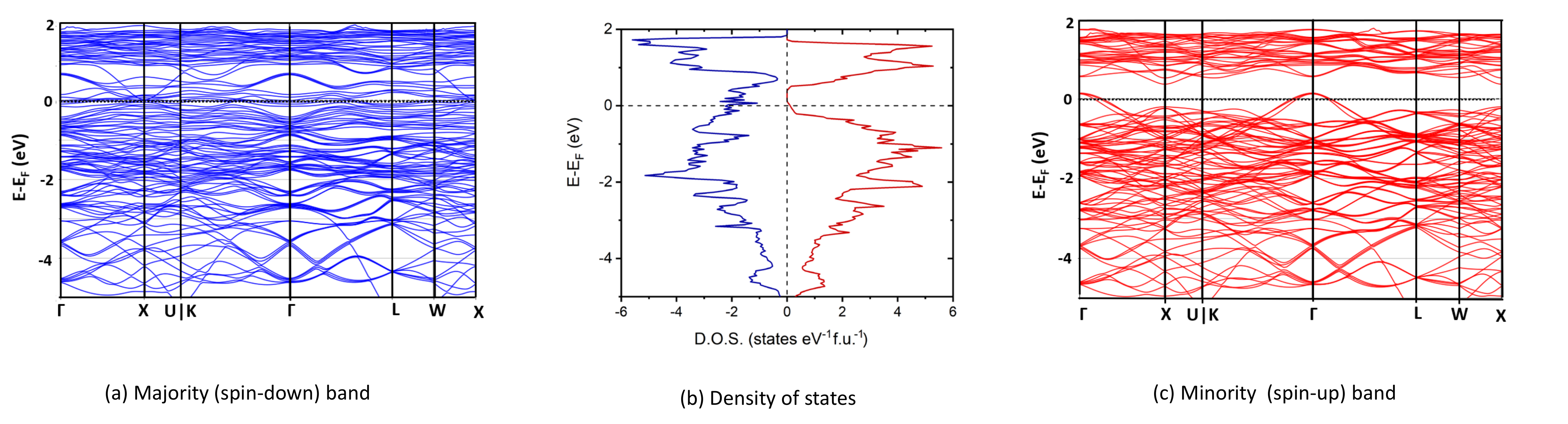}}
{\caption{Spin polarized band structure and density of states of  FeMnVAl in disordered structure: (a) majority (spin-down) band (b) density of states, (c) minority (spin-up) band. The energy axis zero point has been set at the Fermi level, and the spin-up (minority) and spin-down (majority) electrons are represented by positive and negative values of the DOS, respectively.}\label{DOS_Disorder}}
\end{figure*}
As realized from XRD, M\"{o}ssbauer and NMR measurements, a disordered structure with Fe and Mn equally distributed among the 4$c$ and 4$d$ sites is the most likely scenario while considering the crystal structure of FeMnVAl. Consequently, we have revisited the electronic structure analysis by considering such disorder in the system. Expectedly, we find that the enthalpy of formation ($\Delta_f{H}$) for the SQS-disordered structure is smaller than that of the ordered Type-2 structure and estimated to be -34.14\,{\rm kJ/mol} (Table~\ref{Enthalpy}). The lower formation energy ($\Delta$E = -1.44\,{\rm kJ/mol) for the disordered structure, thus confirms that disordered FeMnVAl is energetically more stable than the ordered one, as expected from the M\"{o}ssbauer analysis.

\begin{figure}[h]
\centerline{\includegraphics[width=.48\textwidth]{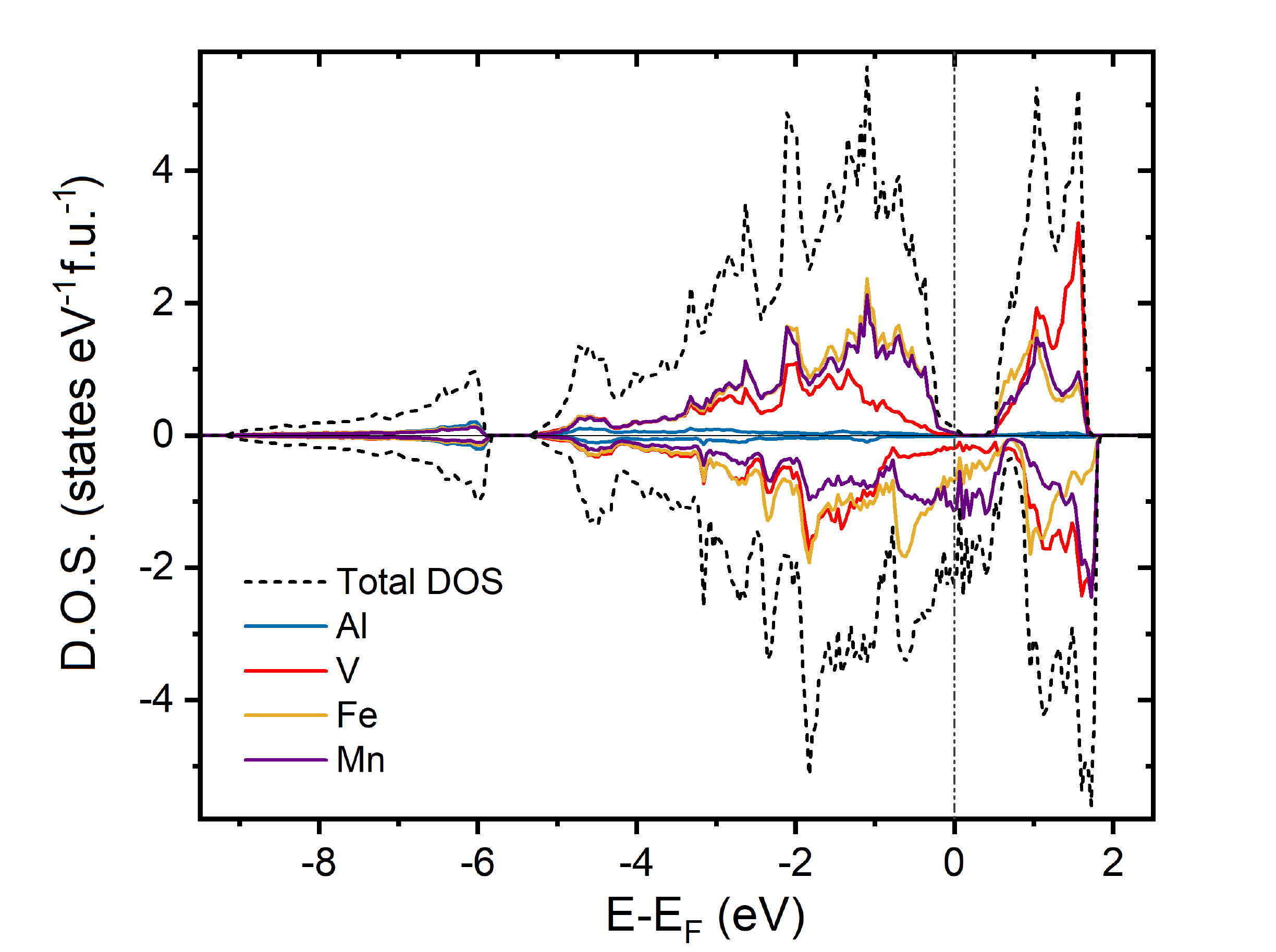}}
{\caption{Electronic DOS (total and partial) of disordered FeMnVAl}\label{Partial_DOS}}
\end{figure}
\begin{figure}[ht]
\centerline{\includegraphics[width=.48\textwidth]{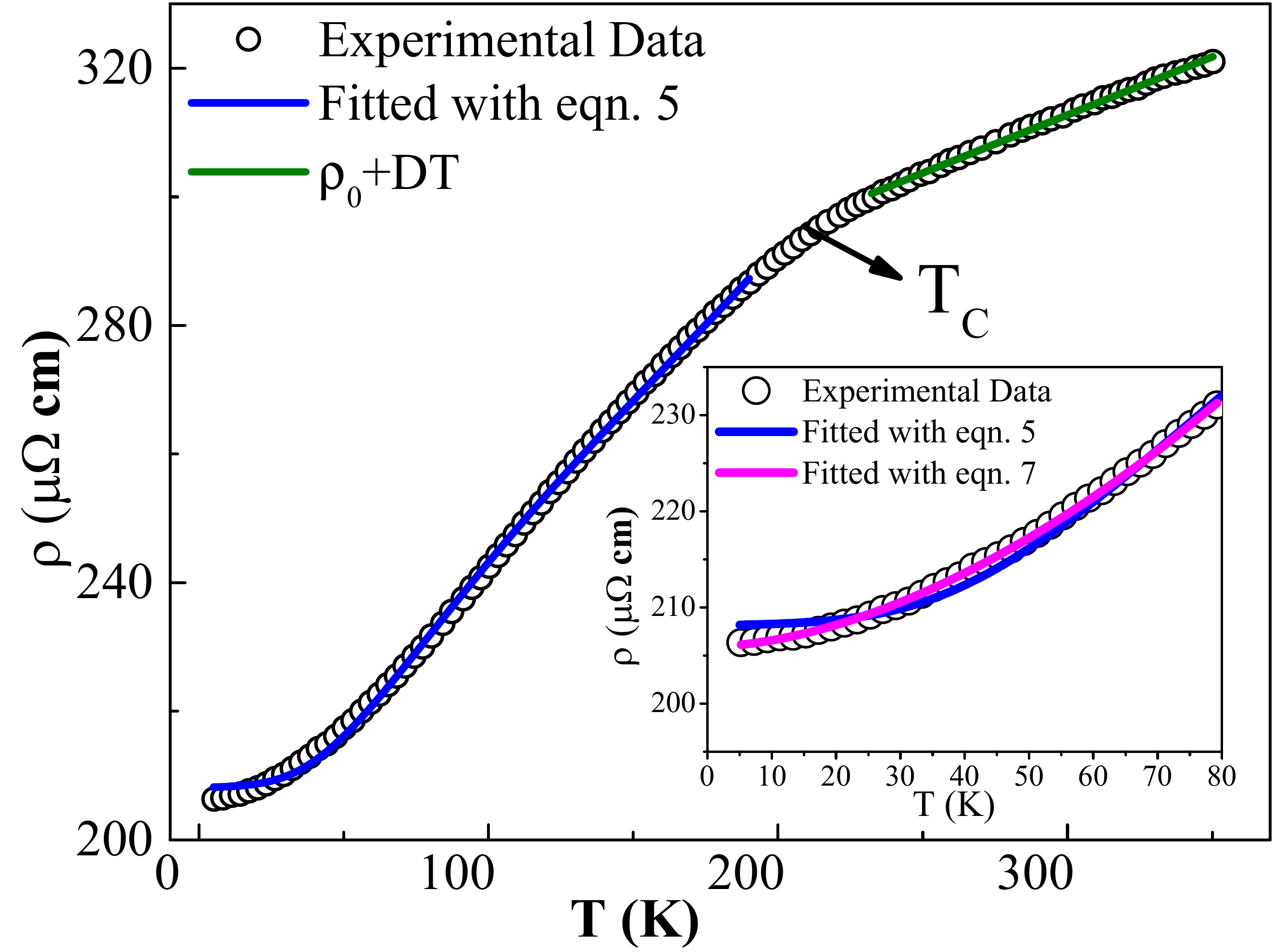}}
{\caption{Temperature dependence of the zero field electrical resistivity in the temperature range 5-350 K.}\label{fig_R_T}}
\end{figure}
The spin-polarized band structure and the density of states for the disordered  structure are presented in Fig. ~\ref{DOS_Disorder}. The Fermi level at the minority (spin-up) band still maintains the band gap implying that disordered FeMnVAl retains its half-metallic ferromagnetic ground state. Despite the disorder, a high spin-polarization is estimated with $P = 90.5 \%$. In fact, the band structure of disordered compounds is on average similar to that of Type-2, despite the local atomic arrangement contributes to decrease the total energy. As in the case of polarization, the total magnetic contribution also remain essentially unchanged \textit{{vis-\`{a}-vis}} the ordered structure with for each element : Fe = -0.27 $\mu_{\rm B}$/f.u., Mn = -1.2 $\mu_{\rm B}$/f.u., V = 0.55 $\mu_{\rm B}$/f.u. and Al = 0.01 $\mu_{\rm B}$/f.u.\@ and presented in Fig.~\ref{Partial_DOS}. The total magnetic moment in disordered structure is thus estimated to be -0.91  $\mu_{\rm B}$/f.u. compared to the order structure (-0.97 $\mu_{\rm B}$/f.u.) and is in agreement with the experimentally observed reduced moment value of 0.84 $\mu_{\rm B}$/f.u. at 70 kOe and 3 K. It can also be noticed that the magnetic contribution of Mn increases significantly at the expense of Fe.

\subsection{\label{sec:Resistivity}Resistivity}

In order to search for the signature of half-metallic ferromagnetism, electrical resistivity of FeMnVAl has been measured both in the absence as well as in the presence of magnetic field  (H = 50 kOe). However, we found negligible changes in resistivity and for clarity, only zero field resistivity data is presented in Fig.~\ref{fig_R_T}. The value of the residual resistivity ratio (RRR), \textit{i.e.}, {{$\rho_{350 {\rm K}}$}}/{{$\rho_{5 {\rm K}}$}}=1.553 is rather low and typical for Heusler alloy. It is known that there are a number of different contributions to the resistivity of a ferromagnetic materials, while Matthiessen rule says that those different scattering mechanisms are independent of each other and additive~\cite{bainsla2016equiatomic}. The total resistivity for  ferromagnetic materials can be written as

\begin{equation}
\rho(T) = \rho_0 + \rho_{P}(T) +\rho_{M}(T)
\label{eqRes1}
\end{equation}
\noindent

where $\rho_0$ is the residual resistivity that originates from the lattice defects, lattice irregularities, \textit{etc.} and the temperature dependent term $\rho_P(T)$ and $\rho_M(T)$ are due to phonon scattering and magnon scattering, respectively.
Phonon scattering term is generally written  as

\begin{equation}
\rho_P = A{\bigg(\frac{T}{\Theta_{D}}\bigg)}^5 \int_{0}^{\frac{\Theta_{D}}{T}} \frac{x^{5}}{(e^{x}-1)(1-e^{-x})} dx
\label{eqRes2}
\end{equation}
\noindent
where \textit{A} is the phonon scattering constant and $\Theta_D$ is the Debye temperature~\cite{gruneisen1933abhangigkeit}. The magnon term which comes from the spin-flip mechanism is quadratic in temperature and can be represented as $\rho_M$ = \textit{B}$T^2$~\cite{bombor2013half}. The $\rho_M$ term persists upto T$_{\rm C}$, but does not make any contribution in the paramagnetic region.

First, we have attempted to fit the whole data below T$_{\rm C}$ using  eqn.~\ref{eqRes1}. The fitted curve (using eqn.~\ref{eqRes1}) describes the experimental data well within the temperature range 80 $<$ T $ <$ 190 K, but the fitted curve fails to trace the experimental data in the low temperature region (5 $<$ T $<$ 80 K). From the fitted parameters, we found that the magnon contribution is very small compared to the phonon contribution. So we have tried to fit the low temperature (5 $<$ T $<$80 K) resistivity data with the equation

\begin{equation}
\rho= B+CT^n
\label{eqRes3}
\end{equation}
\noindent
 which had been utilized in several Heusler-based HMF materials~\cite{bainsla2015corufex, rani2017structural, rani2018origin}. As can been seen in Fig.~\ref{fig_R_T}, the fit to eqn.~\ref{eqRes3} in this temperature range is quite good and  the value of \textit{n} estimated is 1.74. However, this value of \textit{n} $\sim$ 1.74 is not associated with any known kind of  scattering process. It is generally known that when the value of the exponent \textit{n} is not equal to 2, it signifies the absence of magnon contribution. This result is also in consonance with the negligible magnon contribution that we have inferred from the analysis of $\rho(T)$ behavior in temperature range 80$<$ T $<$ 190 K. Here we mention that similar values of the exponent \textit{n} are also reported in literature for different Heusler-based HMF materials. For example,  for CoRhMnGe~\cite{rani2017structural} , the reported value of \textit{n} is 1.53 and  for recently published NiCuFeGa~\cite{samanta2020structural}, the reported value of the exponent is 1.76 which is pretty close to the value obtained in our material. Although the DFT calculation establishes FeMnVAl is a ferromagnetic system, the absence of (or presence of very small) magnetic contribution from the magnon term below T$_{\rm C}$  at first may look surprising. This however could be explained from the fact that in HMFs, one of the sub-bands has negligible DOS at E$_{\rm F}$,  and therefore the magnetic contribution arising from the spin-flip scattering gets considerably diminished. Our resistivity data thus indirectly suggests the presence of the HMF states in FeMnVAl, in agreement with to the DFT calculations.
\\ \\
\section{\label{sec:Conclusion}Conclusion}
A new Fe-based quaternary Heusler alloy FeMnVAl has been synthesized. Theoretical calculation shows that V+Al and Fe+Mn in the same cubic plane (Type-2 ordered structure) has minimum energy and spin-polarized band-structure calculations indicates presence of a half-metallic ferromagnetic ground state. A detailed combined study of XRD and M\"{o}ssbauer spectrometry suggest the presence of site-disorder between Fe and Mn in Type-2 structure, which is also supported by the estimated lower formation energy obtained from theoretical calculations. Magnetic susceptibility exhibits distinct ferromagnetic transition at $T_{\rm C}\sim$ 213 K. ${^{57}}$Fe M\"{o}ssbauer and ${^{27}}$Al \& ${^{51}}$V NMR spectroscopic measurements, coupled with the magnetic susceptibility results, confirms that  Mn is the major contributor of the magnetism which is further supported by first principle calculations. DFT calculations further shows that the value of the spin-polarization changes only nominally from 92.4 \% in  ordered structure to 90.4 \% in disordered structure, which is quite a striking feature and emphasizes the robustness of half-metallicity in FeMnVAl compared to other half-metallic ferromagnets reported in literature. The disordered structure possessing lower formation energy and maintaining high spin-polarization coincide to a very unusual scenario where the introduction of this particular type of disorder actually results in an enhancement of symmetry of the crystal structure from $F\bar{4}3m$ (no. 216) to $Fm\bar{3}m$ (no. 225). Absence (or small contribution) of magnon term in the resistivity data also provides an indirect support towards the presence of a HMF ground state.\\

\section{Acknowledgement}
S.G and S.C would like to sincerely acknowledge SINP, India and UGC, India, respectively, for their fellowship. DFT calculations were performed using HPC resources from GENCI-CINES (Grant 2021-A0100906175). Work at the Ames National Laboratory (in part) was supported by the Department of Energy- Basic Energy Sciences, Materials Sciences and Engineering Division, under Contract No. DE-AC02-07CH11358.
\bibliographystyle{apsrev4-2}

%

\end{document}